\documentclass[sigconf,screen]{acmart}

\AtBeginDocument{%
  }

\copyrightyear{2025}
\acmYear{2025}
\setcopyright{cc}
\setcctype{by-nc-sa}
\acmConference[CHI '25]{CHI Conference on Human Factors in Computing Systems}{April 26-May 1, 2025}{Yokohama, Japan}
\acmBooktitle{CHI Conference on Human Factors in Computing Systems (CHI '25), April 26-May 1, 2025, Yokohama, Japan}\acmDOI{10.1145/3706598.3714104}
\acmISBN{979-8-4007-1394-1/25/04}

\acmConference[Conference acronym 'XX]{Make sure to enter the correct
  conference title from your rights confirmation emai}{June 03--05,
  2018}{Woodstock, NY}

\graphicspath{{figures/}{pictures/}{images/}{./}} 

\usepackage{algorithm}

\usepackage{color}
\usepackage{enumitem}
\usepackage{xspace}
\usepackage{listings}

\newcommand{\ie}{{i.e.,}\xspace}
\newcommand{\eg}{{e.g.,}\xspace}

\newcommand{\etal}{{et~al.}\xspace}

\definecolor{bluecrayola}{rgb}{0.12,0.46,1.0}

\newcommand{\tool}{InterLink\xspace}

\definecolor{mygreen}{RGB}{0,176,80}
\definecolor{mygreen2}{RGB}{232, 245, 233}
\definecolor{myorange}{RGB}{192, 0, 0}
\definecolor{myblue}{RGB}{47, 85, 151}

\newcommand{\yanna}[1]{{\color{black} {#1}}}
\newcommand{\concise}[1]{\textcolor{black}{#1}}

\newcommand{\revise}[1]{{\color{black} {#1}}}

\newcommand{\review}[1]{\textcolor{black}{}}

\newcommand{\final}[1]{{\color{black} {#1}}}

\begin{document}

\title{InterLink: Linking Text with Code and Output in Computational Notebooks}

\lstset{
    basicstyle=\ttfamily\small, 
    keywordstyle=\color{blue}, 
    identifierstyle=\color{black},
    commentstyle=\color{green},
    stringstyle=\color{red},
    frame=single, 
    breaklines=true, 
    showstringspaces=false, 
    tabsize=2, 
    captionpos=b 
}


\author{Yanna Lin}
\authornote{The work was done when Yanna Lin was visiting CMU.}
\orcid{0000-0003-3730-0827}
\affiliation{%
  \institution{The Hong Kong University of Science and Technology}
  \city{Hong Kong SAR}
  \country{China}
}
\affiliation{%
  \institution{Carnegie Mellon University}
  \city{Pittsburgh, PA}
  \country{United States}
}
\email{ylindg@connect.ust.hk}

\author{Leni Yang}
\orcid{0000-0003-4527-4905}
\affiliation{%
  \institution{The Hong Kong University of Science and Technology}
  \city{Hong Kong SAR}
  \country{China}
}
\email{lyangbb@connect.ust.hk}

\author{Haotian Li}
\authornote{The work was done when Haotian Li was at HKUST.}
\orcid{0000-0001-9547-3449}
\affiliation{%
  \institution{Microsoft Research Asia}
  \city{Beijing}
  \country{China}
}
\email{haotian.li@microsoft.com}

\author{Huamin Qu}
\orcid{0000-0002-3344-9694}
\affiliation{%
  \institution{The Hong Kong University of Science and Technology}
  \city{Hong Kong SAR}
  \country{China}
}
\email{huamin@cse.ust.hk}

\author{Dominik Moritz}
\authornote{Dominik Moritz is the corresponding author.}
\orcid{0000-0002-3110-1053}
\affiliation{%
  \institution{Carnegie Mellon University}
  \city{Pittsburgh, PA}
  \country{United States}
}
\email{domoritz@cmu.edu}


\begin{abstract}

Computational notebooks, widely used for ad-hoc analysis and often shared with others, can be difficult to understand because the standard linear layout is not optimized for reading.
In particular, related text, code, and outputs may be spread across the UI making it difficult to draw connections.
In response, we introduce \tool, a plugin designed to present the relationships between text, code, and outputs, thereby making notebooks easier to understand.
In a formative study, we identify pain points and derive design requirements for \final{identifying and navigating} relationships among various pieces of information within notebooks.
Based on these requirements, \tool features a new layout that separates text from code and outputs into two columns.
It uses visual links to signal relationships between text and associated code and outputs and offers interactions \final{for navigating related pieces of information.}
\final{In a user study with 12 participants, those using \tool were 13.6\% more accurate at finding and integrating information from complex analyses in computational notebooks.
These results show the potential of notebook layouts that make them easier to understand.}

\end{abstract}

\begin{teaserfigure}
  \includegraphics[width=\linewidth,alt={Composite figure showing user frustration with a traditional long, linear notebook on the left, and an innovative side-by-side layout of our system on the right. The left side depicts a notebook with dense text and graphics, overlayed with a user's thought bubble expressing difficulty in locating information due to the notebook's complexity. The right side shows our system's interface with observations noted on the left pane and corresponding code and outputs on the right pane, simplifying data navigation and linking results directly to their source code. 
  }]{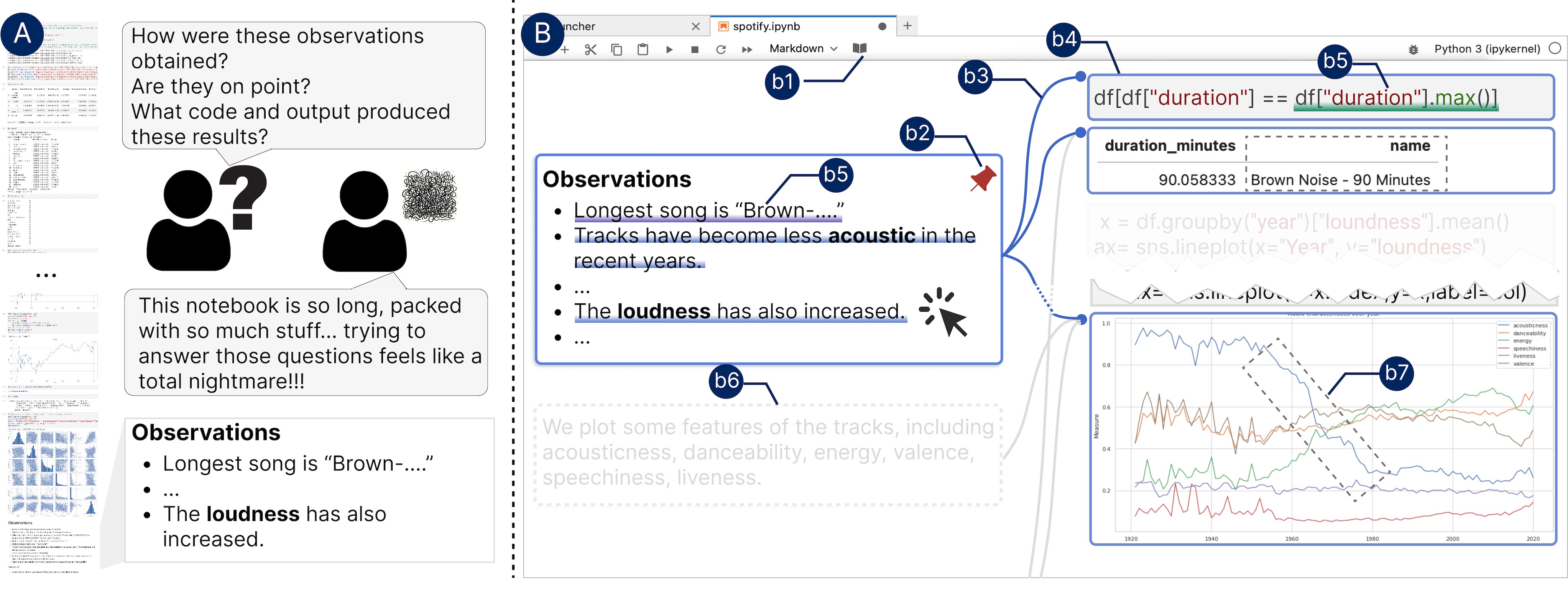}
  \caption{
  \tool helps analysts read and understand computational notebooks by linking text to code and outputs using a side-by-side layout.
  Clicking the icon in the toolbar (b1) activates \tool, switching the layout of the notebook from a linear layout (A) to a side-by-side layout (B).
  In (B), the ``observation'' cell, marked with the fixed icon (b2), indicates that it has been fixed in place through a click, ensuring it remains visible on the screen for easy reference.
  A linking line (b3) explicitly connects the ``observation'' cell to the code cell (b4).
  Visual cues (b5-b7), including colored underlines, dashed borders, and dashed sketches, indicate the existence of relationships between segments or cells.
}
    \Description{Composite figure showing user frustration with a traditional long, linear notebook on the left, and an innovative side-by-side layout of our system on the right. The left side depicts a notebook with dense text and graphics, overlayed with a user's thought bubble expressing difficulty in locating information due to the notebook's complexity. The right side shows our system's interface with observations noted on the left pane and corresponding code and outputs on the right pane, simplifying data navigation and linking results directly to their source code. }
  \label{fig: teaser}
\end{teaserfigure}




\maketitle

\section{Introduction}

Computational notebooks, such as Jupyter~\cite{jupyter} and RStudio~\cite{rstudio}, have become essential tools in data science and research.
Adhering to the literate programming principle, they combine executable \yanna{code}, code outputs, and descriptive text in a single document, making the rationale behind computations accessible~\cite{knuth1984literate}.
These notebooks are structured into discrete units of computation and descriptive text, known as ``cells'', which can be executed and edited independently. 
This flexible design not only streamlines the iterative process of code testing and refinement but also simplifies the integration of comprehensive descriptive text alongside computational content.
As a result, computational notebooks are frequently created and shared among various stakeholders, so there is a wide audience that needs to read and understand the notebooks~\cite{ramasamy2023visualising}.

However, understanding shared computational notebooks is challenging.
Though a notebook author may make the best effort to clean the code and output cells and write descriptive and explanatory text, readers still face three major challenges when \final{trying to identify and navigate the relationships between the text, code, and outputs}\footnote{\final{For brevity, the term ``relationship'' in the remainder of this paper refers to the relationships between the text, code, and outputs.}}.
\yanna{\textbf{The first is to be aware of related text for code or outputs.}}
The location of text relative to its corresponding code or outputs can vary significantly. 
Text describing some code or output can appear above, below, or far away from the relevant code or outputs, depending on the author's preferences or the type of descriptive text~\cite{wang2022documentation}.
Consequently, \yanna{readers may not realize the existence of related text for extra information when reading specific code and output cells, and vice versa}.
\textbf{The second is sorting out the complex relationships between code, outputs, and text, which may exhibit many-to-many relationships at different granularities.}
\yanna{Even when readers are aware that relevant information exists, having a clear view of all the relationships and locating them is challenging.
The relationships between code, outputs, and text often involve many-to-many mappings at different levels of granularity.}
In many-to-many relationships, a single text cell can be related to one or more code or output cells, and several text cells can relate to a single code or output cell. 
Furthermore, these elements can be related at different granularities.
For example, a specific output cell may be related to a segment of a text cell that summarizes findings from multiple output cells.
In such cases, reader\final{s} might find it challenging to \yanna{locate the related content and identify the specific connections} because they need to process and filter extensive information.
\textbf{The third is retaining \yanna{and synthesizing} the relationship information.}
Even when relationships of \yanna{code}, outputs, and text are clarified, readers may still struggle to retain and synthesize them for understanding.
When the notebook is long or \final{contains} unrelated cells interspersed among relevant information, readers have to scroll back and forth to collect, remember, and \final{process} relevant information. 

Existing research on making computational notebooks more readable has predominantly focused on isolated aspects, such as code organization~\cite{head2019managing, weinman2021fork, wang2022stickyland} and descriptive text generation~\cite{lin2023inksight,wang2022documentation}.
These approaches, while valuable, often overlook the interplay between \yanna{code}, outputs, and text.
Recent efforts have aimed to address this gap by implementing structured outlines within computational notebooks~\cite{chattopadhyay2023make, rule2018aiding}. 
Nonetheless, they mainly applied headlines to structure computational notebooks and neglected a substantial portion of the narrative content in other types, which constitutes approximately 68\% of the descriptive text~\cite{wang2022documentation}.
Furthermore, previous studies have reported readers' desire for more granular explanations that extend beyond headers to single code statement explanations~\cite{chattopadhyay2023make}.

\revise{To fill this gap, our work proposes \yanna{a tool}, \tool, that clearly presents relationships \final{between the text, code, and outputs to help readers identify and navigate related information in notebooks, thereby making them easier to understand.}
\final{Rather than specifying these relationships, which we leave for future work, this work investigates whether explicitly presenting these relationships in a reading-optimized design can improve the readability of computational notebooks.}
\yanna{To inform \tool's design, we first conducted a formative study involving semi-structured interviews with six participants.
The study revealed challenges in \final{identifying and navigating} the relationships between text, \yanna{code}, and outputs.
Based on these findings, we derived five key design requirements to guide the design of ~\tool.
}
\yanna{Before designing \tool to present relationships, we first designed a relationship space to identify the types of relationships to present and proposed their corresponding formulation.}
This space defines 27 distinct relationship types, considering both content type (\yanna{code}, outputs, and text) and their granularity (ranging from a whole cell to specific segments). 
It includes relationships such as a text segment describing ``the song with \final{the} longest duration'' and a corresponding visual output segment indicating ``the tallest bar in a bar chart''.
Based on user feedback from the formative study that emphasized the importance of the text-code and text-output relationships,
we developed \tool, a plugin to \final{make notebooks easier to understand} by clarifying these relationships. 
}
Specifically, \tool separates text from \yanna{code} and outputs by presenting them in two columns, aligning related descriptive text side by side with its corresponding computational elements for immediate reference, as illustrated in~\autoref{fig: teaser} (B). 
It uses explicit connection lines for cell-cell connections and subtle highlights for intricate, fine-grained relationships. 
\revise{\tool~also provides interactions for \final{navigating} relationships.}
To demonstrate the usefulness and effectiveness of \tool, we conducted a user study with 12 participants.
\yanna{
Our study shows that \tool improves task accuracy \final{by 13.6\% over the traditional notebook interface} \final{in identifying, navigating, and integrating} relevant information within computational notebooks.
}
Finally, we discussed future directions for facilitating the understanding of notebooks.
\revise{The main contributions of this paper are as follows:
\begin{itemize}
\item 
A formative study to identify the readers' pain points and derive design requirements of \final{identifying and navigating the relationships between text, code, and outputs in computational notebooks;}

\item
\yanna{A novel computational notebook plugin}, \tool, to present relationships between descriptive text and associated \yanna{code} and outputs for facilitating the understanding of notebooks;
\item
A user study to demonstrate the usefulness and effectiveness of \tool.
\end{itemize}
}

\section{Related Work}
Our research is related to prior studies on how people understand computational notebooks, computational notebook layouts, and \yanna{links between code, outputs, and text.}

\subsection{Computational Notebook Understanding}
\label{sec:rw_understanding}

\yanna{Existing tools communicate analysis results in computational notebook\final{s} by transforming them} into other formats such as slides~\cite{zheng2022nb2slides, li2023notable, wang2024outlinespark, wang2023slide4n} and data comics~\cite{kang2021toonnote}. 
\yanna{While these tools focus on transforming notebook\final{s into other formats}, our work emphasizes \final{reading and} understanding computational notebooks in their original form to support continuous analysis and reuse.}

\yanna{
The iterative nature of exploratory data analysis often leads to long, messy notebooks, compounded by insufficient or absent textual explanations, which hinder reading, understanding, and collaboration~\cite{rule2018aiding}.
To tackle these issues, 
some research has explored textual explanation generation methods to improve readability and understanding~\cite{lin2023inksight, wang2022documentation, wang2020callisto}. 
Other efforts focus on enhancing code organization to mitigate messiness, including code cleaning and organization~\cite{head2019managing, wang2020assessing, shankar2022bolt}, adapting notebooks for non-linear analysis workflows~\cite{weinman2021fork, wang2022stickyland, harden2022exploring}, and providing analytical workflow overviews~\cite{ramasamy2023visualising, wenskovitch2019albireo}.
}
Despite these advancements, these approaches typically focus on a single aspect—either \yanna{code} or text—without fully addressing the complex interplay between them, a critical component for reading and understanding notebooks~\cite{kery2018story}. 

To fill this gap, recent research has introduced structured outlines to better organize the loose \yanna{code} and text. 
For instance, Rule \etal~\cite{rule2018aiding} developed a plugin that allows users to group code cells with annotated text, incorporating features like text cell folding for detailed exploration. 
\yanna{Similarly, Chattopadhyay \etal~\cite{chattopadhyay2023make} organized notebooks into labeled, expandable chapters for improved navigation.}
Though useful, they predominantly emphasize headers, overlooking a significant portion of descriptive content—approximately 68\%—that exists in forms other than headers \cite{wang2022documentation}. 
Users have indicated a need for understanding content \final{at a finer level of granularity}, such as individual code statements over broader, header-level summaries~\cite{chattopadhyay2023make}. 
Aligning with these observations, our work aims to address this gap by providing \revise{an interface} that facilitates a clearer understanding of multi-granular relationships between text, code and output.

\subsection{Computational Notebook Layout}
\label{sec:layout}

Computational notebooks have become essential in data science and computational research, improving the way computational narratives are constructed and presented~\cite{rule2018exploration}.
Yet, the prevailing linear structure of these notebooks often poses challenges for non-linear analyses~\cite{wang2022stickyland, weinman2021fork}, effective navigation through extensive content~\cite{harden2022exploring}, and the orderly management of complex information~\cite{head2019managing, liu2019understanding, shankar2022bolt}, ultimately affecting usability and comprehension.

Researchers have explored various strategies to address these limitations by redesigning the notebook layout to leverage large displays and support complex analytical tasks.
For example, Wang \etal proposed StickyLand~\cite{wang2022stickyland}, a plugin that treats each cell as a movable sticky note, allowing for a flexible,
non-linear organization of \yanna{code} that aligns with the exploratory nature of data analysis. 
Weinman~\etal~\cite{weinman2021fork} introduced forking mechanisms within notebooks facilitating the exploration of multiple analytical pathways.
Additionally, empirical studies further showed the benefits of multi-column or workboard layouts of notebook\final{s} to branching and comparative analyses~\cite{harden2022exploring, harden2023there}.
Besides non-linear analysis, B2~\cite{wu2020b2} offers interactive visualizations alongside notebook content, aiming to bridge the gap between the iterative, two-dimensional nature of interactive visualizations and the linear layout of computational notebooks. 
Janus~\cite{rule2018aiding} further enhances notebook functionality by introducing collapsible content sections, which can be selectively displayed to minimize clutter and focus on relevant analyses.


Observable reflected on the traditional notebook single-column, narrow layout for presentation purposes and noted its limited information density.
\revise{They pointed out that while this format is suitable for ad-hoc exploration, it is inadequate for presentations and displays~\cite{observable}.}
Thus, a different representation of notebooks, optimized for reading and understanding, would be valuable.
In light of these developments, our work proposes a side-by-side layout with interactions and visual cues for computational notebook\final{s}, aiming to enhance information integration and understanding by presenting the relationships between text with code and output.

\subsection{\final{Code-text and Code-output Links}}

\yanna{

Computational notebooks are widely used in data science and research because they integrate executable code, outputs, and descriptive text into a cohesive document, making the rationale behind computations more transparent~\cite{knuth1984literate, wang2022documentation}.
However, synthesizing different information often challenging especially when they are spatially separated~\cite{kong2014extracting}.  

To address this issue, research in programming environments has explored linking code with relevant texts or outputs to streamline tasks like coding and debugging. 
Some tools enhance coding workflows by linking code to related texts, such as interactive documentation~\cite{oney2012codelets, ko2006barista}, chat messages~\cite{oney2018creating}, or code explanations~\cite{canvas, yan2024ivie}. 
These tools often present information inline, adjacent, or in floating panels for efficient information seeking and understanding.
While effective, these approaches typically focus on simple, one-to-one relationships between code and text.
Others connect source code of GUI application\final{s} with corresponding UI outputs to support code review and debugging~\cite{huang2024unfold, chi2018doppio}.
Extending beyond one-to-one relationships, UNFOLD~\cite{huang2024unfold} labels code pieces with multiple UI output identifiers, yet it still requires users to piece together the relationships manually.

Recent tools, which extend computational notebook interface for additional content creation, provide interactions for users to link back to the original content in computational notebooks.
For instance, when using OutlineSpark~\cite{wang2024outlinespark} to create slides next to a computational notebook interface, users can click the slide outline to highlight related notebook cells.
Similarly, Callisto~\cite{shi2020calliope} uses a two-column layout to link chat messages with specific code or outputs.
However, these tools require users to actively trigger connections and lack mechanisms to provide an overview of relationships across text, code, and outputs within a notebook.

To address these gaps, our work introduces \tool, a notebook plugin designed to present complex many-to-many relationships between code, outputs, and text.
\final{\tool reorganizes code, outputs, and text by computing layouts based on their relationships, making it easier for users to cross-reference information.
The layout is further enhanced by line connections and visual cues, allowing users to see the relationships, thereby facilitating locating and integrating relevant information.}}


\subsection{\final{Text-visual Links}}
\yanna{
Visual elements, including charts and images, are often closely connected with descriptive text, and linking them is critical for effective data communication and comprehension~\cite{badam2019elastic, lalle2021gaze, zhi2019linking}. 
Prior research has extensively explored the value of explicit connections between text and visuals to improve readability.
For example, 
Zhi~\etal validated that connecting text with visualizations significantly reduces user reading time while improving comprehension task performance~\cite{zhi2019linking}. 
Thus, several authoring tools
have been developed to help authors link text with visual elements~\cite{WonderFlow,dataplaywright, sultanum2021leveraging}.
These tools use video or scrollytelling~\cite{seyser2018scrollytelling} to present text-visual links. 
However, these methods rely on a fixed content exploration order determined by the author, which limits the reader's ability to explore the content flexibly according to their individual needs.

Active reading systems have attempted to address this limitation by enabling readers to organize or interact with linked information. 
For example, LiquidText allows readers to freely select and organize content on a workspace through multitouch interactions~\cite{tashman2011liquidtext}.
While effective, such an approach sacrifices the original document structure, which is essential in computational notebooks where computation order conveys important contextual meaning.
Other tools have supported in-situ information presentation to enhance reading through context-aware linking~\cite{badam2019elastic, lalle2021gaze, zhi2019linking}.
For instance, Badam~\etal developed a system that allows users to select text sentences and link them to corresponding visual elements in tables or visualizations using synchronized visual highlighting~\cite{badam2019elastic}. 
However, these methods primarily support on-demand, localized comprehension while failing to provide a global overview of how different content components are interconnected.

While previous work shows that linking text with visuals eases reading, existing methods cannot be directly applied to computational notebooks.
To address this gap, we propose \tool, a notebook plugin that links text with code and outputs \final{to make notebooks easier to understand}. 
\tool visualizes relationships by using connecting lines that provide a global cell-cell relationship overview and subtle highlights for fine-grained relationship details, while preserving the computation flow of the notebook.
It also offers interactions for exploring these relationships flexibly.
}

\section{Formative Study} \label{sec: formative study}

We conducted a formative study to capture the challenges of understanding computational notebooks and to derive design requirements for facilitating the understanding of notebooks. 
This study involved semi-structured interviews with six participants who have experience working with shared computational notebooks.

\subsection{Participants}
We recruited six participants (3 females, 3 males; aged $26.8\pm2.2$ years; identified as FP1-6) through online advertisements on social media and word-of-mouth referrals.
All participants are postgraduate students, with five pursuing PhD degrees. 
They have extensive experience using Python and Jupyter Notebooks, \final{with an average of} $6.8\pm1.9$ years and $5.3\pm0.8$ years, respectively. 
On a 5-point Likert scale (where 1 indicates ``not at all familiar'' and 5 indicates ``extremely familiar'')~\cite{likert}, all participants rated themselves as either ``moderately familiar'' (4) or ``extremely familiar'' (5) with Python and Jupyter Notebooks. 
Their backgrounds span diverse fields, including data visualization, computer vision, natural language processing, reinforcement learning, and human-computer interaction.

\subsection{Procedure}

The formative study was executed through one-on-one online meetings. 
Participants began by signing a consent form that authorized use to record videos of the session and to collect their demographic information and feedback for research purposes.

\concise{We first collected participants' familiarity with Python and Jupyter Notebooks and their demographic information.
After that, participants were asked to describe their experience with understanding computational notebooks.
They were asked to share recent notebooks they had attempted to understand to provide a concrete context and details of the challenges encountered.}
\concise{When} sharing notebooks was impractical due to data confidentiality, participants were asked to select a notebook of interest from Kaggle competitions using their domain expertise as relevant keywords (\eg~``computer vision'').
Participants were then asked to thoroughly comprehend the selected or shared notebooks until they felt confident enough to reuse the notebooks or for up to 20 minutes (limited to prevent fatigue). 
Then we asked the participants to describe the challenges they encountered in understanding the notebook.
We \concise{asked} open-ended questions such as, ``Did you encounter any challenges while understanding this notebook?'', ``What aspects, if any, hindered your understanding of this notebook?'', ``Have you faced similar challenges in previous experiences with notebooks?'', and ``Beyond this notebook, have you encountered other obstacles that made understanding notebooks difficult?''.
\concise{We further asked participants to provide detailed explanations and concrete examples with follow-up questions like, ``Can you elaborate on why this aspect hinders your understanding?'' and ``Could you provide specific examples?''}
%
Finally, we invited the participants to propose what tool features they expected to enhance the understanding process.
Each session lasted around an hour, with participants compensated with US \$15 for their time and insights.

\subsection{{Findings}}
\label{sec:findings}


\yanna{Participants in \final{the} formative study reported both obstacles in understanding individual elements of notebooks (i.e., text, code, and outputs), as well as the structure and relationships of these elements.
While previous studies (\autoref{sec:rw_understanding}) have predominantly discussed and addressed problems related to individual elements and structured outlines, this section reports obstacles caused by the complex relationships and interleaving structure of these elements.}

\textbf{\yanna{C1: Difficult relationship inference due to} unclear, multi-granular relationships between code, outputs, and text.}
Participants identified challenges in inferring the relationships among text, code, and outputs (FP1-FP4).
They mentioned spending extra time determining the source or reference point of the text within the notebook.
These challenges were notably intensified when attempting to associate specific segments across these elements.
For instance,
FP1, who often reviews notebooks for research purposes, noticed a trend where notebooks use markdown cells with summaries to describe findings from previous analyses. 
Yet, he struggled to trace back each finding to its corresponding output and encountered even greater difficulty when attempting to pinpoint specific segments within those outputs.
Similarly, when delving into code and outputs, participants needed additional efforts to identify related textual descriptions for understanding (FP1, FP3). 

\yanna{\textbf{C2: Demanding cross-reference due to single-column layout.}
Participants reported that the traditional single-column layout made it difficult to cross-reference related information, requiring frequent scrolling to check information across different notebook cells (FP1, FP2, FP4). 
FP4 mentioned, ``\textit{When the code or text extends beyond my screen, I have to scroll back and forth to check details, which is mentally demanding as I have to remember and recall them.}''
FP1 echoed this concern, adding that the challenge intensifies when related cells are far apart.}

\textbf{C3: Disrupted focus due to interleaving structure.}
Participants in our study (FP1, FP2, FP5) reported that the interleaving structure of notebook elements often disrupted their focus, \yanna{making it difficult to maintain sustained attention on specific type\final{s} of elements}.
FP1 noted a feeling of interference when attempting to concentrate on a single aspect of the notebook. 
\yanna{Similarly, FP2 described difficulty focusing on text during the understanding phase and code during the applying phase~\cite{forehand2010bloom}, due to the interleaving text and code}.
FP5 mentioned that displaying various types of information (including text, \yanna{code}, and outputs) in limited vertical screen space is \yanna{dizzying} and overwhelming, \yanna{making it hard to focus on one type of element at a time.}

\subsection{Design \yanna{Requirements}}

\yanna{Findings in~\autoref{sec:findings} reveal that readers encounter numerous obstacles in \final{identifying and navigating} the relationships among text, code, and outputs.}
\yanna{Based on these findings, we derive five key design requirements for our system design.}

\yanna{\textbf{DR1: Provide clear visualization with clutter-reducing mechanisms for multi-granular relationship inference.}}
To address the difficulties in inferring the unclear and multi-granular relationships between \yanna{code}, outputs, and text, the tool should visually show these relationships \textbf{(C1)}.
These visual aids should support the full spectrum of relationship granularities, ranging from entire cells to specific segments within them.
\yanna{Moreover, the tool should incorporate mechanisms to minimize visual clutter caused by numerous relationships.}
With the tool, readers should be able to effortlessly infer both the existence and positions of the \yanna{relationships}.

\yanna{\textbf{DR2: Offer flexible interactions to streamline relationship inference and navigation.}}
The tool should offer flexible interactions to allow users to \yanna{easily infer and \final{navigate} relationships and cross-reference the connected information \textbf{(C1 and C2)}. 
With the tool, readers should be able to easily locate, navigate, and cross-reference interconnected text, code, and outputs} at various granularities dispersed across the notebook, thereby facilitating the synthesis and interpretation of these interconnected elements.

\yanna{\textbf{DR3: Provide alternative layouts for efficient cross-referencing.}
The tool should explore alternatives to the single-column layout, organizing related information in a more accessible manner for quick cross-referencing \textbf{(C2)}.
With the tool, users should be able to easily infer multi-granular relationships and cross-reference related information for better understanding.
}

\textbf{DR4: Enable separate focus on text or code and output \yanna{to reduce focus disruption}.}
The narrative text and computational \yanna{code} or outputs are often intertwined, which can disrupt user focus when attempting to concentrate on one type of element.
To address this, the tool should provide mechanisms to isolate the content of the text with \yanna{code} and outputs \textbf{(C3)}.
With the tools, readers should be able to selectively concentrate on either the text or the \yanna{code} and outputs independently. 

\textbf{DR5: Integrate with existing platforms seamlessly.} \yanna{Supported by literature showing that integrating with existing platforms reduces users' learning curve~\cite{li2023notable}, we propose this design requirement.}
The tool should integrate seamlessly with common computational notebook environments, eliminating the need for readers to learn a new interface.

\revise{In this research, we aim to understand whether a reading-optimized design can improve the experience of reading computational notebooks. We leave authoring tools for future work.} 

\section{\tool}

\autoref{sec:relationship_space} introduces relationship spaces, outlining the types of relationships based on content--text, \yanna{code}, and outputs--and granularity, from whole cells to segments.
\autoref{sec:relationship_formulation} then  describes how to formulate the relationships. 
Finally, \autoref{sec:system_design} details the design of \tool, starting with an overview of the system, followed by the layout computation, relationship visualization, and interaction.

\begin{figure}[t!]
    \centering
    \includegraphics[width=\linewidth,alt={Diagram illustrating the relationship space of a tool, with circles representing segments and rectangles representing entire cells. The top part shows inter-category relationships among code, text, and output, connected by solid and dashed lines with 8 key relationships highlighted in red to represent the primary focus of the tool. The bottom part demonstrates intra-category relationships within the same content type, depicted with dashed black lines. A legend on the right details the symbols and line styles used to indicate cell-cell, cell-segment, and segment-segment relationships.}]{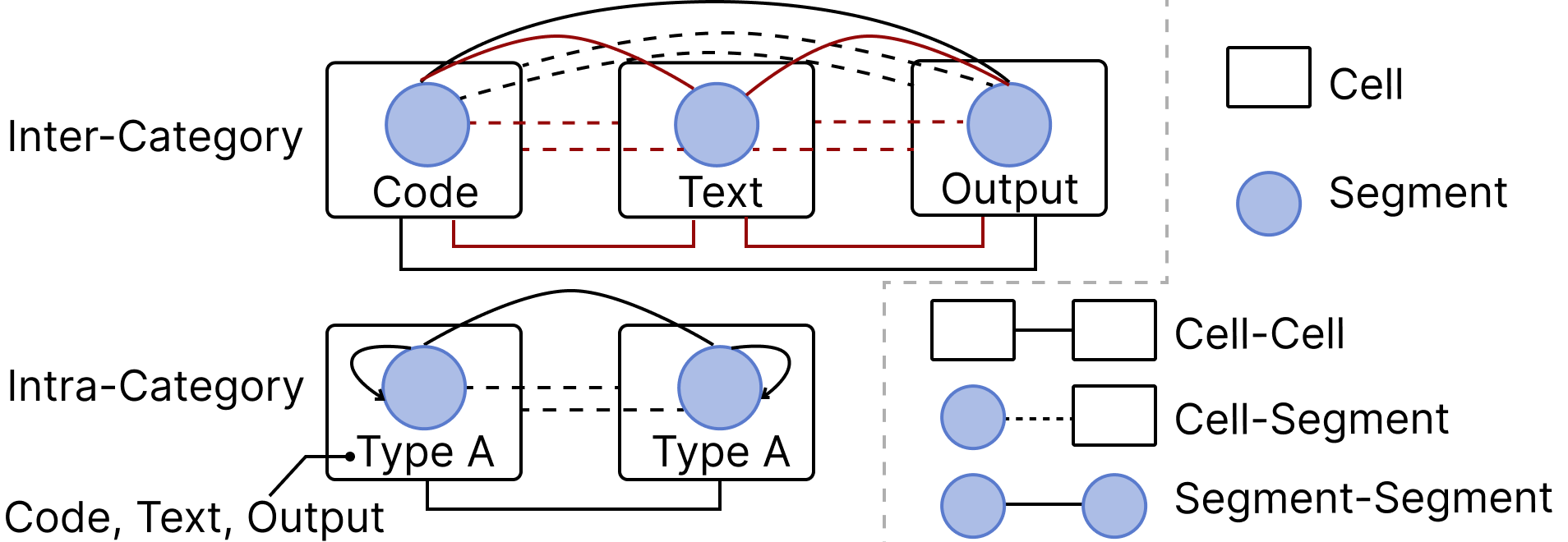}
    \caption{The relationship space in computational notebooks, categorized by content types (\ie~text, code, and output) and granularity (entire cell and segments).
    It identifies 27 distinct relationship types: 12 stemming from inter-category and 15 from intra-category relationships.
    Specifically, 8 inter-category relationship types highlighted in \textcolor[rgb]{0.796,0.255,0.329}{red} are the primary focus of \tool.}
    \Description{Diagram illustrating the relationship space of a tool, with circles representing segments and rectangles representing entire cells. The top part shows inter-category relationships among code, text, and output, connected by solid and dashed lines with 8 key relationships highlighted in red to represent the primary focus of the tool. The bottom part demonstrates intra-category relationships within the same content type, depicted with dashed black lines. A legend on the right details the symbols and line styles used to indicate cell-cell, cell-segment, and segment-segment relationships.}
    \label{fig:relationship}
    \vspace{-1em}
\end{figure}

\subsection{Relationship Space}
\label{sec:relationship_space}

Considering both the content categories (including text, code, and output) and the granularity of the cells (including the whole cell and \final{segments} of cells), we build a complete relationship space between all of them, as shown in~\autoref{fig:relationship}.
Regarding content, relationships within notebooks can be categorized into both intra-category and inter-category relationships: 1) Intra-category relationships occur within the same category of notebook content (\ie~text-text, code-code, and output-output relationships);
and 2) Inter-category relationships span different categories of content (\ie~text-code, text-output, and code-output relationships).
Additionally, we defined the granularity of these relationships, identifying cell-cell relationships that exist between entire cells, and further nuanced relationships including cell-segment and segment-segment relationships. 
Through this classification, we have identified a total of 27 unique relationship types—12 stemming from inter-category and 15 from intra-category relationships.
There are more intra-category relationships because there are more segment-segment relationships within the same cell category,
\yanna{\eg~multiple code segments within a single code cell contributing to a shared function.}

Feedback from participants in the formative study mainly emphasized challenges associated with text-code and text-output relationship\final{s}.
Thus, in this paper, our focus narrows to these two kinds of relationships, which include eight relationship types (marked in red in \autoref{fig:relationship}).

\subsection{Relationship Formulation}
\label{sec:relationship_formulation}

\begin{figure*}[t!]
    \centering    \includegraphics[width=\textwidth,alt={xx}]{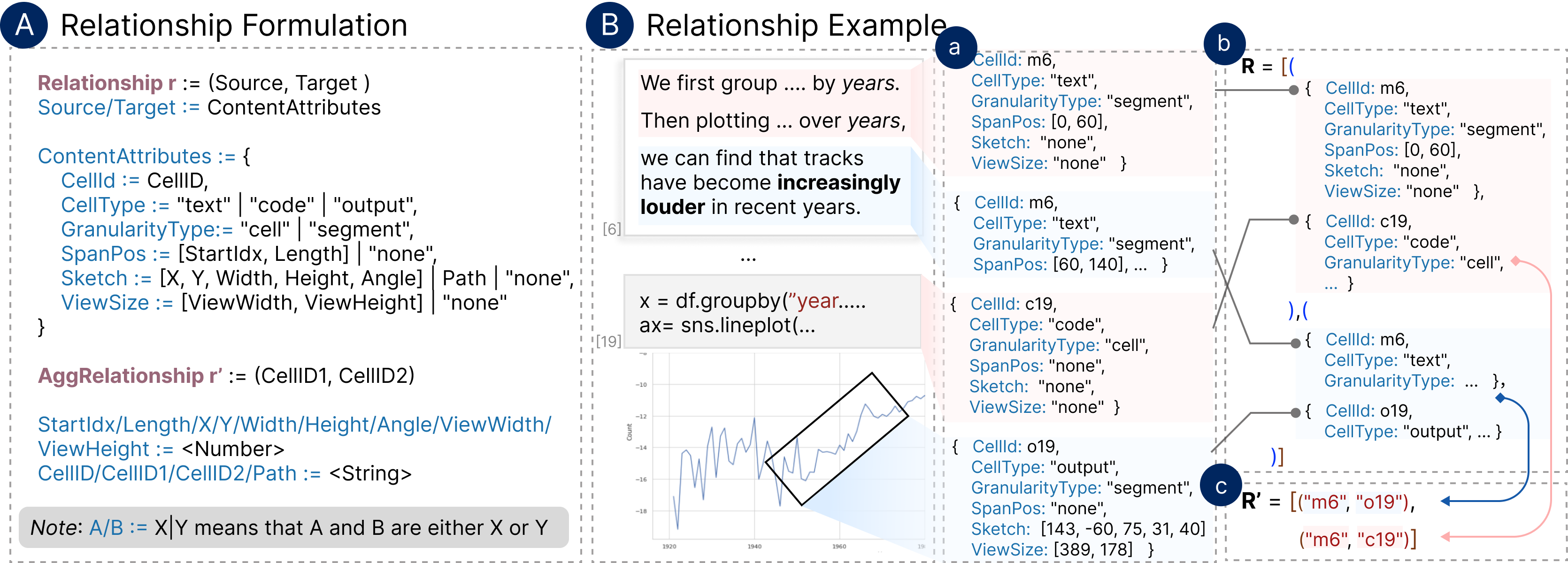}
    \caption{\yanna{This figure shows the relationship formulation and its example.
    (A) describes the formulation of a relationship and the corresponding aggregated relationship. 
    (B) provides an example of how relationships are represented.  (a) shows four examples of how the components in a relationship are represented according to the defined formulation. 
    (b) shows a relationship set $R$, consisting of two relationships formed from these four components. (c) shows the corresponding aggregated relationship set $R'$. }}
    \Description{This figure illustrates the formulation and representation of relationships in computational notebooks, divided into two panels: (A) Relationship Formulation and (B) Relationship Example. Panel (A) defines the structure of a relationship, which consists of a source and target, both represented as ContentAttributes. These attributes include the CellId, a unique identifier for the cell; the CellType, which specifies whether the cell is a "text," "code," or "output" cell; the GranularityType, which indicates whether the relationship applies to the entire cell or a specific segment; and the SpanPos, which identifies the start index and length for locating text or code content, or is marked as "none" if not applicable. For output annotations, the Sketch attribute accommodates either bounding boxes defined as [X, Y, Width, Height, Angle] or freehand sketches recorded as paths, with "none" used if no sketch is provided. The ViewSize attribute records the dimensions of the SVG associated with the annotation as [ViewWidth, ViewHeight], or is marked as "none" if not applicable. Below the main formulation, the aggregated relationship, AggRelationship, is defined as a higher-level summary that uses only the CellId of the two connected cells. Panel (B) provides an example of how these relationships are instantiated and aggregated in a notebook.  Subfigure (a) of (B) displays four examples of how the components in a relationship are represented according to the defined formulation.  Subfigure (b) shows how these four components form two distinct relationships, collectively constituting a relationship set $R$. The first relationship links two segments of a markdown cell, identified as m6, to a code cell, identified as c19. The second relationship links the same markdown cell to an output cell, identified as o19. For each component, attributes such as CellId, CellType, GranularityType, and SpanPos are specified. Subfigure (c) illustrates the corresponding aggregated relationship set $R'$, which simplifies the representation by focusing on the CellId of the two connected cells. In this example, $R'$ consists of two pairs:{"m6", "c19"} and {"m6", "o19"}, summarizing the high-level connections between the notebook cell.}
    \label{fig:rm_json}
    \vspace{-1em}
\end{figure*}

\yanna{ 
We propose a formal specification for a relationship $r$ between text, code, and outputs, as illustrated in~\autoref{fig:rm_json} (A).
Each relationship consists of two components, each defined by several attributes. 
The \textit{CellId} and \textit{CellType} identify the ID and type of the cell involved, while \textit{GranularityType} specifies whether the relationship applies to the entire cell or specific segments. 
For text or code content, \textit{SpanPos} precisely locates the relevant segment by specifying its starting index and length.
For output annotations, \textit{Sketch} accommodates the multi-modal nature of outputs (\eg~images, text, and tables) by supporting rectangular bounding boxes or freehand sketches recorded as paths. 
Additionally, \textit{ViewSize} captures the SVG dimensions of the annotations to ensure precise positioning across varying screen sizes.
To provide an overview of relationships between notebook cells without specific details, we further introduce an aggregated relationship $r'$ derived from $r$, represented by the IDs of the two involved cells.
The relationships in a notebook are collectively represented as $R = [r_1, r_2, \dots]$ and the corresponding aggregated relationships as  $R' = [r'_1, r'_2, \dots]$.

\autoref{fig:rm_json}~(B) illustrates an example of how relationships are represented. (a) shows four examples of how the components in relationships are represented according to the defined formulation.
These components form a relationship set $R$ with two distinct relationships, linking two text segments to one code cell and one output segment, respectively (b). 
The corresponding aggregated relationship set $R'$ is shown in (c).

}

\begin{figure}[t!]
    \centering
    \includegraphics[width=\linewidth, alt={The figure is a screenshot of the InterLink interface. On the left side, there is a series of text cells, with one labeled ``Observations'' providing a list of insights about Spotify's song data. On the right side, the code and output cells are displayed. The code cells include Python code, utilizing the seaborn library to create visualizations of the data. The output cell shows a line plot titled ``Count of Tracks Added'', displaying the number of tracks added over time. In the middle, connecting lines visually link the corresponding text cells with their related code or output cells, highlighting the relationships between explanations and their data-driven visualizations.}]{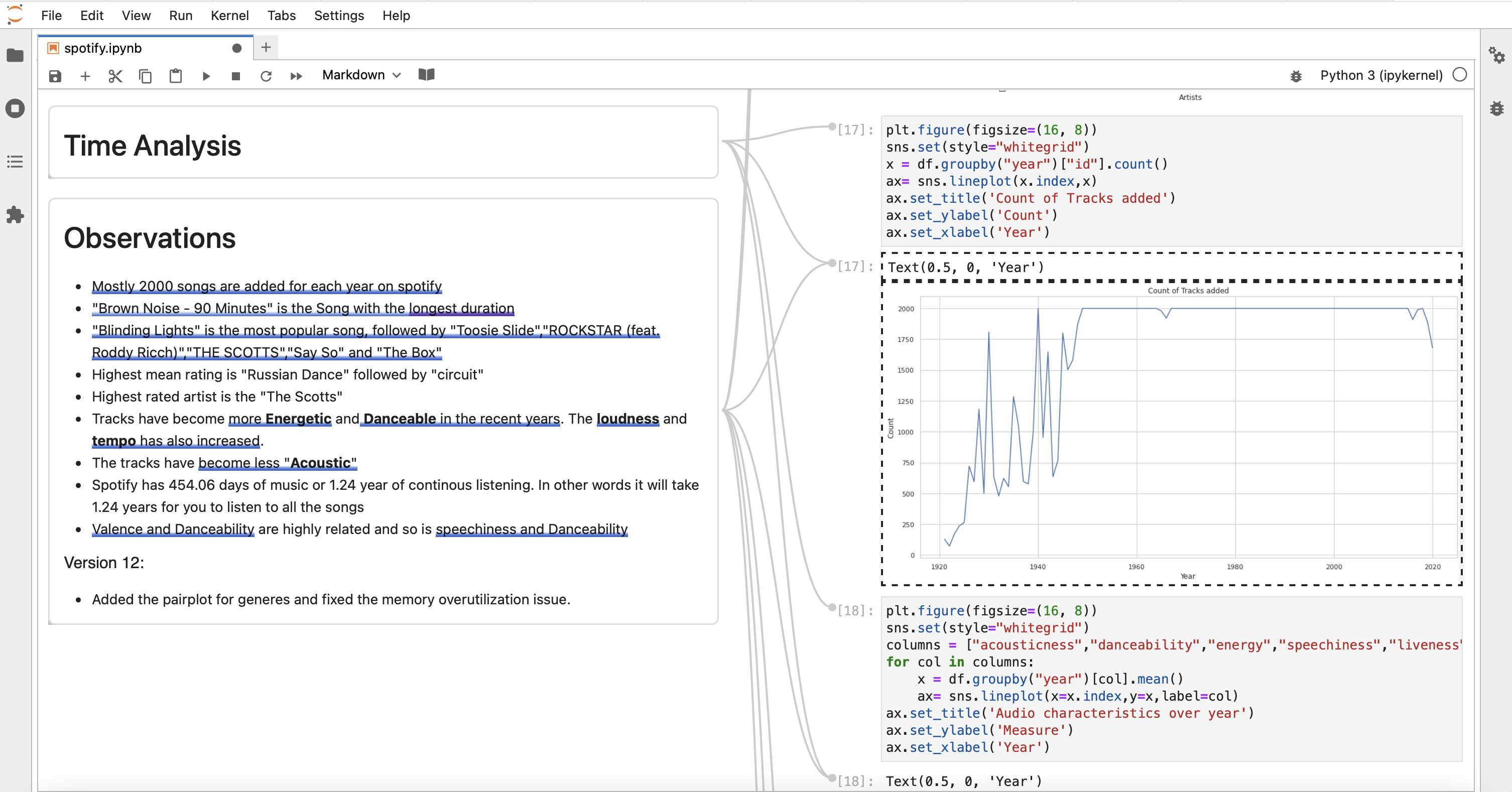}
    \caption{\revise{Screenshot of the \tool interface, with text cells on the left, code and output cells on the right, and lines in the middle denoting aggregated relationships between them.}}
    \Description{The figure is a screenshot of the InterLink interface. On the left side, there is a series of text cells, with one labeled ``Observations'' providing a list of insights about Spotify's song data. On the right side, the code and output cells are displayed. The code cells include Python code, utilizing the seaborn library to create visualizations of the data. The output cell shows a line plot titled ``Count of Tracks Added'', displaying the number of tracks added over time. In the middle, connecting lines visually link the corresponding text cells with their related code or output cells, highlighting the relationships between explanations and their data-driven visualizations.}
\label{fig:interlink_interface}
\end{figure}

\subsection{System Design}
\label{sec:system_design}
In this section, we first present an overview of \tool.
Then we introduce the details of the layout computation, relationship visualization, and supported user interactions.

\subsubsection{System Overview}
\label{sec:overview}
We developed \tool, a plugin for JupyterLab. \tool equips readers with visualizations and interactive features that facilitate \final{identifying and navigating} relationships between text with \yanna{code} and outputs (\textbf{DR5}), thereby making it easier to \final{understand} notebooks \revise{(\autoref{fig:interlink_interface}).}

When readers launch \tool in a notebook by clicking a button in the toolbar (\autoref{fig: teaser}~(b1)), the notebook is shown in two side-by-side columns: the left column dedicated to the text cells, and the right column for code and output cells (\textbf{DR3, DR4}),
\revise{as illustrated in~\autoref{fig: teaser}~(B).
\autoref{fig:interlink_interface} provides an actual screenshot of the interface.}
Between the two distinct columns, there are lines connecting text cells with their related code or output cells (\textbf{DR1}), delineating the presence of relationships at a cell level without specifying the details within~(\autoref{fig: teaser}~(b3)).
This design choice emphasizes connections between whole cells without the granularity of segment-specific relationships, aiming to reduce visual clutter (\textbf{DR1}).
To detail the nuanced, multi-granular information underpinning these relationships, individual cells have visual cues (\textbf{DR1}).
Specifically, color-coded underlines within text and code cells draw attention to specific text segments involved in a relationship (\autoref{fig: teaser}~(b5)), while dashed sketches over the output cell emphasize particular segments of interest (\autoref{fig: teaser}~(b7)). 
Additionally, a dashed border surrounding an entire cell signals that the relationship encompasses the whole cell (\autoref{fig: teaser}~(b6)).

Exploration of the relationships is facilitated through interaction (\textbf{DR2}). 
Key interactions include hovering over elements to immediately highlight related content, which simultaneously invokes tooltips for rapid insights (\autoref{fig:focus_mode} (a1-a5)). 
A ``Shift'' key-triggered focus mode allows for an in-depth examination of relationships, effectively minimizing distractions from unrelated information (\autoref{fig:focus_mode}). 
Additionally, readers can click to fix the position of selected cells, anchoring their focus for detailed analysis (\autoref{fig: teaser}~(b2)).

\subsubsection{Layout Computation} \label{sec:layout_computation}



We use \final{a} \yanna{heuristic-based approach guided by} a set of objectives to strategically arrange the cells in a side-by-side layout. In this layout, text cells are on the left, and code and output cells are on the right, as shown in~\autoref{fig:interlink_interface}.
Next, we explain how \tool handles the positions and sizes of cells in the layout.
In terms of positions, \tool retains the primary order of code and output cells. 
For text cells, \tool repositions them to \yanna{follow the original order and place them as close as possible to the first related code or output cell.}
This is informed by~\cite{wang2022documentation}, which suggests that the positions of text cells relative to related code cells are more flexible (\eg~above, below, or adjacent).
In terms of sizes, \tool limits the maximum height of text cells while maintaining the original size of code and output cells.
This decision is informed by the observation that the content length of text cells can be approximately ten times that of code cells~\cite{wang2022documentation}, causing significant height discrepancies. 
In a side-by-side layout with reduced cell width\final{s}, these differences are magnified, leading to misalignment and excessive scrolling similar to single-column layouts.
\yanna{If the content of a text cell exceeds this maximum height, users can access the overflow through manual scrolling or auto-scroll triggered by hovering over its related information.}


Specifically, the following objectives are applied:
\begin{itemize}
    \item[O1] Maintain the original textual narrative order and computational order.
    \item[O2] Be close to the first related code or output cells. This close arrangement aims to enable users to quickly associate textual explanations with relevant computational content and facilitate easy reference. 
    \yanna{\item[O3] Follow the original sequence in the notebook if the text cell is not related to any code or output cells.}
    \item[O4] Adjust the height of text cells according to the cumulative height of \final{the} related cells.  \yanna{\tool assumes that the importance of a text cell is positively correlated with the amount of its related computational content~\cite{lin2023dashboard}. \tool sets the maximum height of a text cell to the cumulative height of its associated code or output cells.}
    \item[O5] Adjust the height of text cells to improve space utilization. 
    \yanna{If a text cell is unrelated to any code or output cells and is followed by a code cell without accompanying text (leaving space to the left of the code cell), its height matches that of the code cell. 
    Otherwise, \tool adds space within the code sequence to fit the text cell with a default height, sacrificing some space but ensuring alignment between text and associated code and output.}
    \item[O6] Ensure that the height of text cells does not exceed the height required to display their content. 
\end{itemize}

The highest priority is given to objective O1, with the others being of equal priority. A detailed pseudocode can be found in the supplementary material.

\autoref{fig:layout_computation} demonstrates the implementation results.
In terms of positions, the textual narrative order on the left and the computational order on the right follow\final{s} the original order strictly (meeting O1).
Notably,  $m1$ is placed near $c1$ to align with O2 for easy referencing.
Meanwhile, $m2$ also tries to approach $c1$ but is placed below $m1$ to maintain the original order, prioritizing order over proximity (O1 over O2).
Moreover, the position\final{s} of $c2, m3, c3$ and $m4$ maintain the original sequence (satisfying O3).
In terms of sizes, the height of $m_1$ equals the sum of the heights of $c_1$ and $o_1$ (meeting O4), while the height of $m_3$ mirrors that of $c_3$ to maintain the original sequence without introducing additional spaces (meeting O4 and O6). Moreover, $m_4$ is set to a default height, and additional spaces are introduced in the computational sequence since the next code cell $c_4$ has relationships with $m_5$ (meeting O5 and O6).

\begin{figure}[t!]
    \centering
    \includegraphics[width=\linewidth, alt={The figure is divided into two parts: the top part illustrates the original notebook and its re-layout after calculating the position and size of each text cell, with corresponding objectives, while the bottom part explains these calculations and objectives in detail. In the top part, the left side shows the original arrangement of the text (m1-m5) and code/output cells (c1-c4 and o1). The re-layout section shows how the text cells are repositioned, with additional space added to align text cell m5 with code cell c4, ensuring computational order and visual clarity. In the bottom section, the left side outlines the formulas for calculating the position and height of text cells based on relationships with code/output cells. On the right, two categories of objectives are listed: "Objectives - Position" (O1-O3) focuses on maintaining narrative and computational order, proximity to related cells, and respecting original order for unrelated cells; "Objectives - Size" (O4-O6) aims to adjust the height of text cells for better space utilization while ensuring content is fully displayed.}]{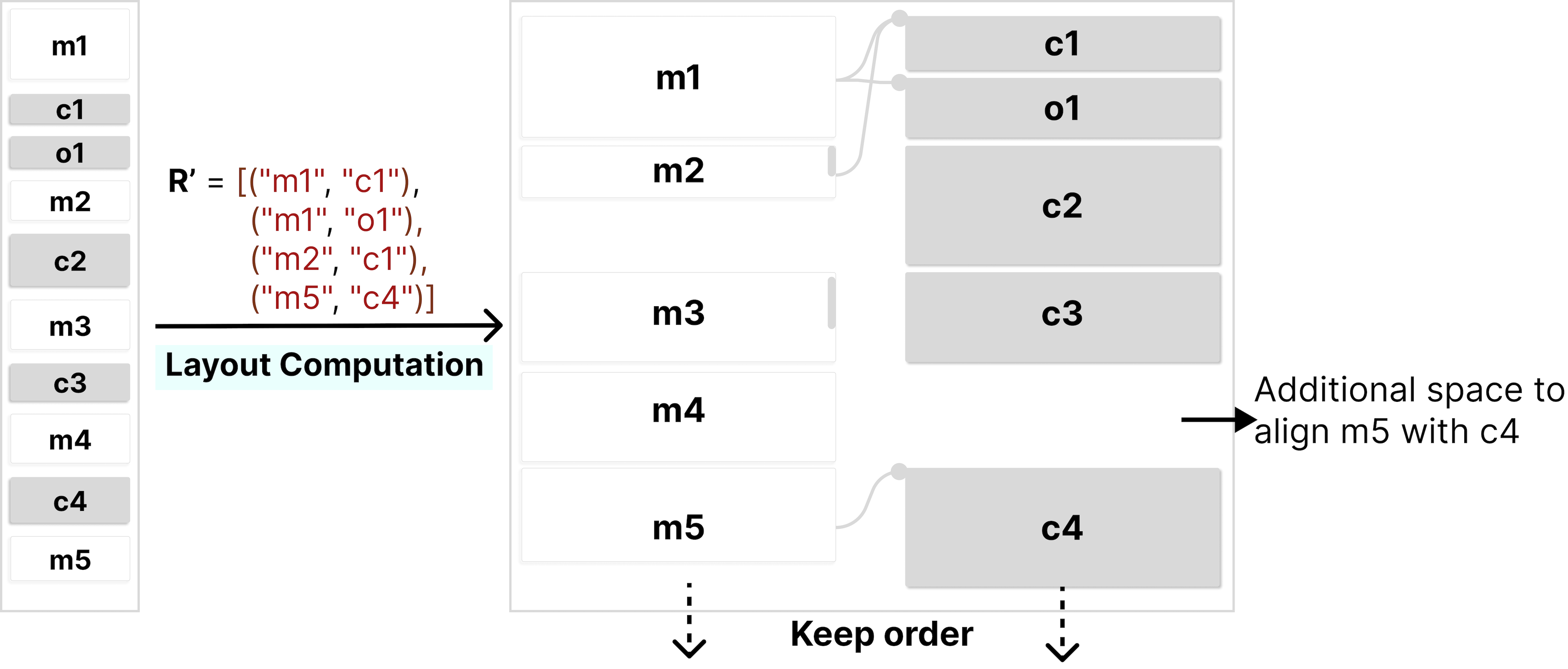}
    \caption{\revise{An example of repositioning and resizing text cells while respecting their original order and relationships with associated code and output cells, as described by the six objectives in \autoref{sec:layout_computation}.}}
    \Description{The figure is divided into two parts: the top part illustrates the original notebook and its re-layout after calculating the position and size of each text cell, with corresponding objectives, while the bottom part explains these calculations and objectives in detail. In the top part, the left side shows the original arrangement of the text (m1-m5) and code/output cells (c1-c4 and o1). The re-layout section shows how the text cells are repositioned, with additional space added to align text cell m5 with code cell c4, ensuring computational order and visual clarity. In the bottom section, the left side outlines the formulas for calculating the position and height of text cells based on relationships with code/output cells. On the right, two categories of objectives are listed: "Objectives - Position" (O1-O3) focuses on maintaining narrative and computational order, proximity to related cells, and respecting original order for unrelated cells; "Objectives - Size" (O4-O6) aims to adjust the height of text cells for better space utilization while ensuring content is fully displayed.}
\label{fig:layout_computation}
\end{figure}

\subsubsection{Relationship Visualization}
\label{sec:relationship_visualization}

Besides structuring notebooks side-by-side, \tool shows relationships in $R$ in the UI.

\begin{figure}[t!]
    \centering    \includegraphics[width=\linewidth,alt={This figure presents a table by \tool that demonstrates visual cues for varying granularity levels, content types, and interaction statuses within a computational notebook. The 'Entire Cell' row indicates inactive state with a dashed border, while the active state is shown with a pink background and a solid blue border. In the 'Segment' row, the content type differentiation is depicted: text segments have underlines in green for code-related, blue for output-related, and purple for both. Their active states are shown with a pink background. The 'Code' segment is underlined in green, which changes to a pink background when active. For 'Output', a dashed black border highlights the area of interest, which changes to a solid red border when active.}]{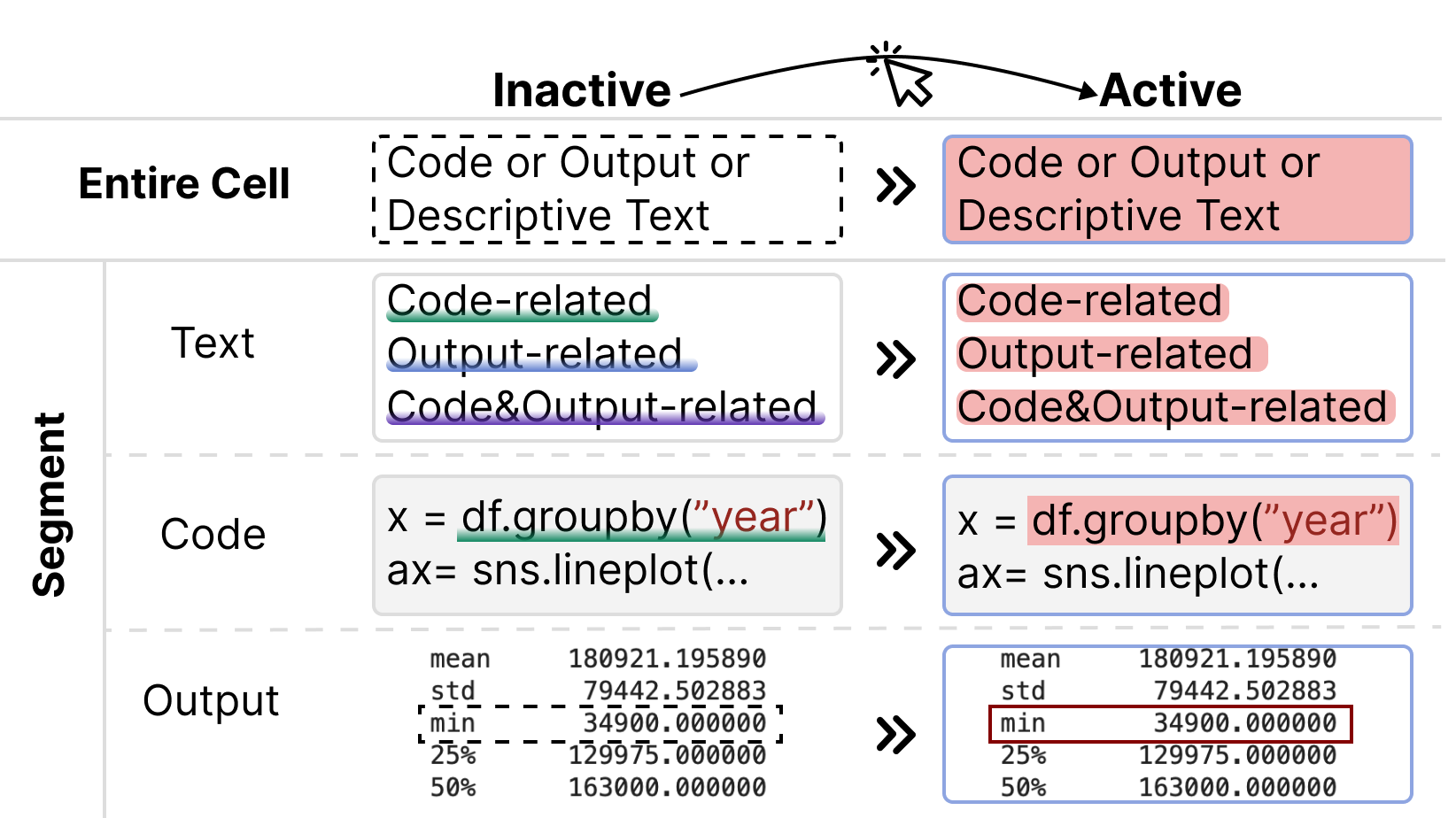}
    
    \caption{Demonstration of how \tool uses visual cues to indicate relationships within the notebook, tailored specifically to the content type, granularity, and interaction status.}
\Description{This figure presents a table by \tool that demonstrates visual cues for varying granularity levels, content types, and interaction statuses within a computational notebook. The 'Entire Cell' row indicates inactive state with a dashed border, while the active state is shown with a pink background and a solid blue border. In the 'Segment' row, the content type differentiation is depicted: text segments have underlines in green for code-related, blue for output-related, and purple for both. Their active states are shown with a pink background. The 'Code' segment is underlined in green, which changes to a pink background when active. For 'Output', a dashed black border highlights the area of interest, which changes to a solid red border when active.}

    \label{fig:visual_cues}
    \vspace{-1em}
\end{figure}

To help readers infer potential relationships within notebooks, \tool adopts visual cues to signal the existence of \yanna{relationships}. \revise{The designs are inspired by~\cite{Srinivasan2019voder}, which summarizes embellishment options for highlighting visualizations to aid interpretation}.
\autoref{fig:visual_cues} shows visual cues, tailored to content type, granularity, and interaction status.
For entire cells, \tool uses dashed borders to signify \final{that} they share a relationship with other information within the notebook.
For segments, the visual cues become more nuanced.
Within text cells, color-coded underlines distinguish the text types:  blue for code-related text, green for output-related text, and purple for text relating to both \yanna{code} and outputs. 
Code segments within code cells use the same green underlines as the code-related text for emphasis.
Specific areas within outputs are outlined with sketches in dashed borders to signify relevant segments, \revise{catering to \final{their} multi-modal nature (\eg~images, tables, and texts).}




To effectively show the detailed relationships identified in $R$, \tool uses a two-tiered approach for visualizing connections \revise{following the ``overview first, then details on demand''~\cite{shneiderman2003eyes}.
Specifically, \tool adopts \textbf{explicit links} to present the high-level aggregated relationships $R'$, providing readers with an overview of the connected information. Additionally, it offers \textbf{implicit interactions} for the granular correspondences within $R$, enabling readers to explore the details as needed. }
We initially considered using explicit lines for all relationships in $R$.
However, this approach causes visual clutter due to the large number of relationships, and content being hidden by links related to detailed segments. 
To mitigate these issues, \tool opts for a more nuanced visual strategy. 
\tool adopts explicit links to visualize the aggregated relationships $R'$, thus offering an overview of the interconnections between text, code, and output cells. 
For instance, \autoref{fig: teaser}~(B) depicts how text cells are interconnected with code and output cells via lines, signifying their relational bonds. 
Specifically, (b3) is the line that depicts the ``observation'' cell related to the code cell (b4).
\revise{Although visualizing the aggregated relationship in $R'$ instead of $R$ can decrease the number of lines displayed, thus reducing visual clutter,  this method may obscure the specifics of the relationships.}
\revise{To address this problem, \tool introduces interactions to facilitate exploring and understanding detailed relationships as needed, which is detailed in~\autoref{interaction_design}.}

\begin{figure*}[t!]
    \centering
    \includegraphics[width=1.0\textwidth, alt={The figure contains four sub-images, illustrating different interaction modes in a re-layout computational notebook. Top-left image: The original re-layout of the computational notebook. The notebook contains multiple cells, including text cells and code/output cells. Lines visually connect related content between text cells (left) and code/output cells (right), showing how content and code are linked. Top-right image (A): Depicts the "hover-to-highlight" interaction. When a user hovers over a specific part of a text cell, a blue outline highlights the hovered text cell itself and the related code or output cell on the right, along with the detailed contents in pink background. Bottom-left image (B): Shows the "key-activated focus" interaction. When activated, unrelated cells are filtered out, leaving only the related cells in screen. This mode helps the user concentrate on relevant content by minimizing distractions from other cells. Bottom-right image (C): Represents the "click-to-fix" interaction. This allows the user to click on a specific cell to pin it in place. A red pin icon appears in the corner of the cell, visually indicating that the cell is now fixed, while the rest of the content can be scrolled independently. This disrupts the default narrative flow but provides easy access for referencing fixed cells.}]{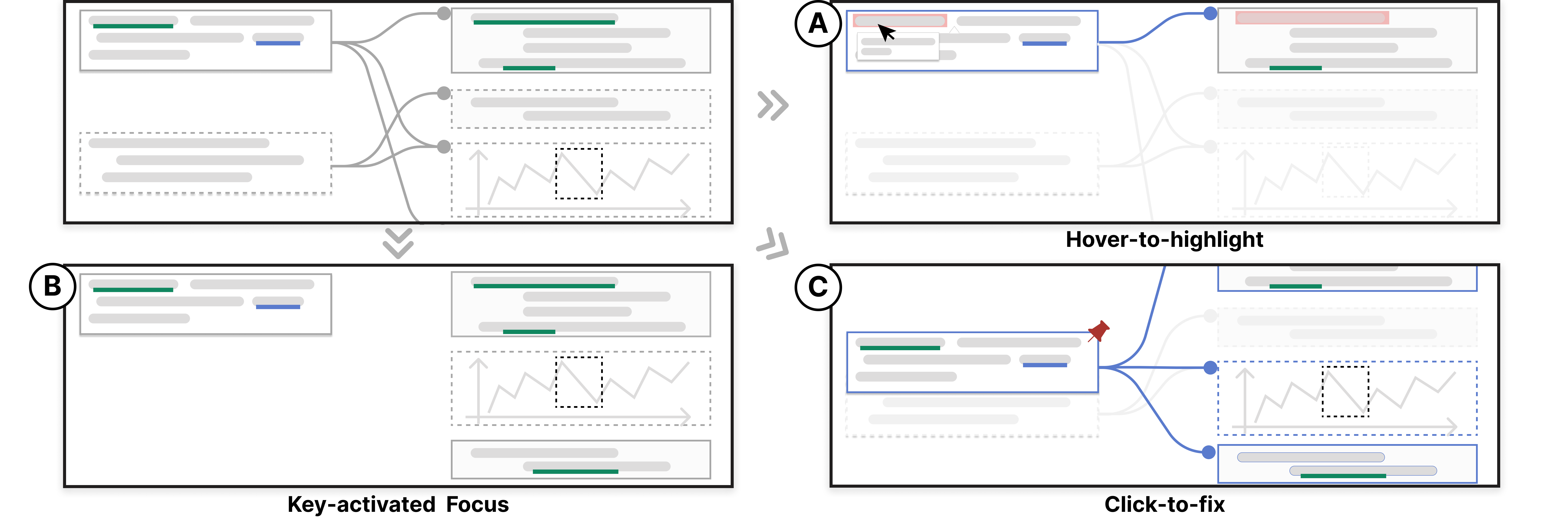}
    \caption{Illustration of three interactions supported by \tool: (A) hover-to-highlight, which aids in identifying detailed corresponding content; (B) key-activated focus, which enables users to concentrate on relevant information while minimizing distractions from unrelated content; and (C) click-to-fix, which allows users to pin a cell, thereby disrupting the original narrative flow to facilitate easy referencing.}
    \Description{The figure contains four sub-images, illustrating different interaction modes in a re-layout computational notebook. Top-left image: The original re-layout of the computational notebook. The notebook contains multiple cells, including text cells and code/output cells. Lines visually connect related content between text cells (left) and code/output cells (right), showing how content and code are linked. Top-right image (A): Depicts the "hover-to-highlight" interaction. When a user hovers over a specific part of a text cell, a blue outline highlights the hovered text cell itself and the related code or output cell on the right, along with the detailed contents in pink background. Bottom-left image (B): Shows the "key-activated focus" interaction. When activated, unrelated cells are filtered out, leaving only the related cells in screen. This mode helps the user concentrate on relevant content by minimizing distractions from other cells. Bottom-right image (C): Represents the "click-to-fix" interaction. This allows the user to click on a specific cell to pin it in place. A red pin icon appears in the corner of the cell, visually indicating that the cell is now fixed, while the rest of the content can be scrolled independently. This disrupts the default narrative flow but provides easy access for referencing fixed cells.}
\label{fig:interaction_space}
\end{figure*}

\subsubsection{Interaction}
\label{interaction_design}

\tool is designed to make it easier to \final{identify} and navigate these relationships, as illustrated in ~\autoref{fig:interaction_space}. 
\revise{These interactions include hover-to-highlight (A), key-activated focus (B), and click-to-fix interactions (C).}

\begin{figure}[t!]
    \centering    \includegraphics[width=\linewidth,alt={This figure demonstrates the `focus mode' of an `observation' cell in a computational notebook interface, with the observation cell on the left and the related information composed on the right, devoid of irrelevant details. The visualization includes labels a1 through a5, which signify a bidirectional hovering mechanism. a3 and a5 are two tooltips appearing to offer succinct explanations.}]{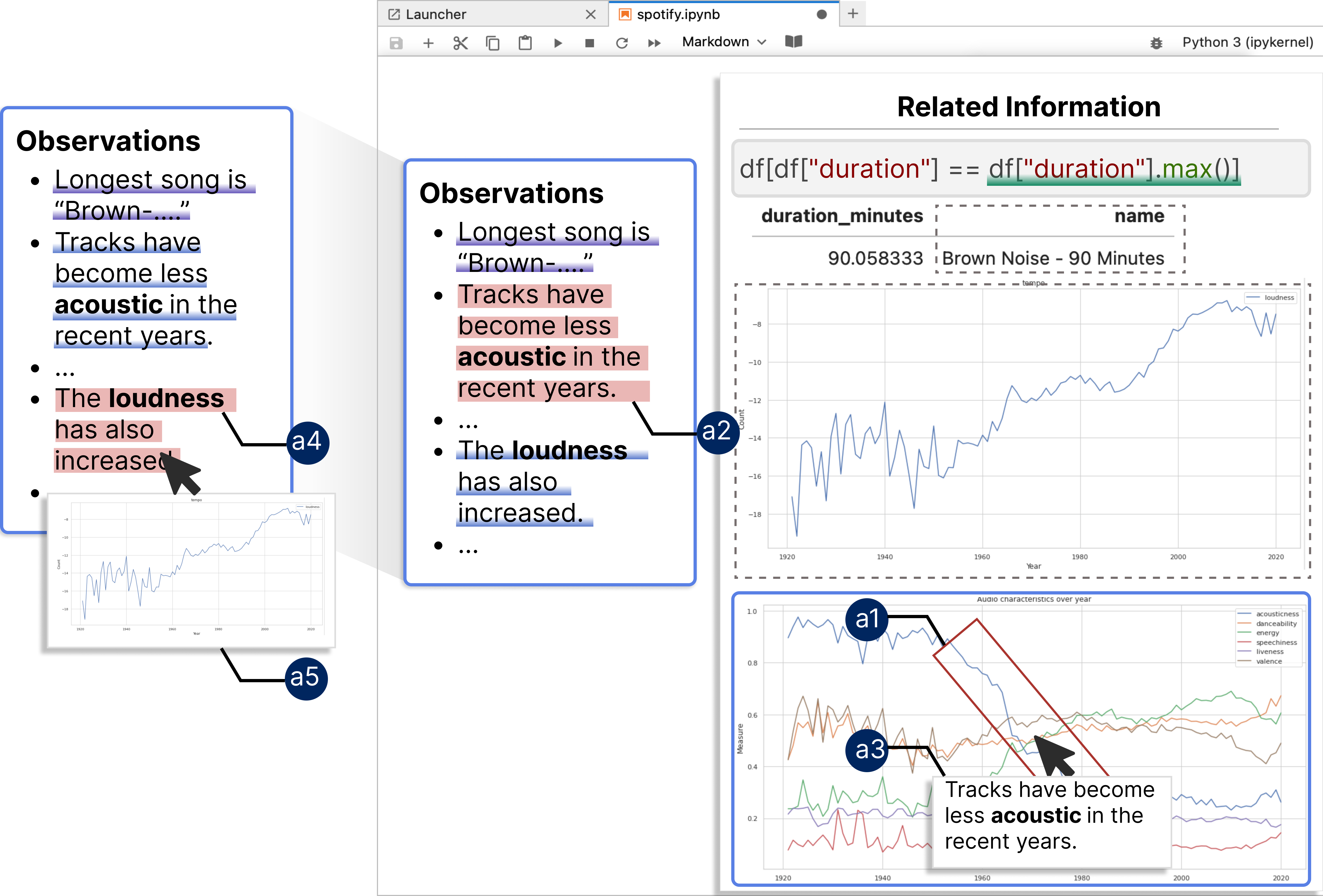}
    \caption{
    The focus mode of ``observation'' cell,  displaying only relevant information and omitting unrelated content. \yanna{
    Hovering over an output segment (a1) highlights the related text in pink (a2) and triggers a tooltip for quick information access (a3). Similarly, hovering over a text segment (a4) triggers a corresponding output snapshot in a tooltip (a5).
    } 
    }
    \label{fig:focus_mode}
    \Description{This figure demonstrates the `focus mode' of an `observation' cell in a computational notebook interface, with the observation cell on the left and the related information composed on the right, devoid of irrelevant details. The visualization includes labels a1 through a5, which signify a bidirectional hovering mechanism. a3 and a5 are two tooltips appearing to offer succinct explanations.}
    \vspace{-1em}
\end{figure}

\textbf{Hover-to-highlight.}
Interacting with lines, cells, or specific visual cues within the cells triggers a highlighting effect.
The borders of related cells and their interconnecting lines turn blue. 
Visual cues related to the hovered content are also marked as ``active'' status with pink backgrounds or red borders, as detailed in~\autoref{fig:visual_cues}. 
This interaction is further complemented by a tooltip that presents concise, relevant content for a quick overview. 
For instance, as shown in \autoref{fig:focus_mode}, hovering over an output segment (a1) changes the background color of the associated content (a2) to pink and brings up a tooltip (a3) next to it for immediate insight. 
Similarly, hovering over a text segment (a4) displays relevant output details in a tooltip (a5).
This feature helps readers identify the relationships.

\revise{\textbf{``Shift'' key-activated focus.} Considering that related cells might be scattered or distant from each other, interleaved with many unrelated ones across the UI, \tool introduces a ``Shift'' key-activated focus mode to address these challenges.}
The selection of the ``Shift'' key as the activation mechanism was deliberate, chosen to avoid conflicts with JupyterLab's existing keyboard shortcuts, such as double-clicking for edit mode, right-clicking for contextual menus, and pressing \final{the} ``Space'' key for scrolling.
\revise{Focus mode filters out all unrelated cells, allowing users to concentrate solely on the information related to the cell or visual cue under examination.
This mode aims to streamline the exploration process within the notebook by minimizing distractions from irrelevant content.}
\autoref{fig:focus_mode} displays the focus mode for the ``observation'' text cell that contrasts with the standard right column interface seen in \autoref{fig: teaser} (B). 
In this mode, only information related to the ``observation'' cell is shown and presented concisely. 

\textbf{Click-to-fix. }\revise{To further address the challenge of related cells not being on the same screen—thus eliminating the need for users to frequently scroll back and forth—\tool supports click-to-fix interactions that allow readers to pin a cell. }
This action locks the cell's position on the screen and maintains its highlight, facilitating a focused examination of its related content without the necessity to adhere strictly to the notebook's linear narrative.
To aid in locating related cells more easily, \tool scrolls automatically upon fixing a cell. 
Hovering over any visual cue automatically brings the first related cell into view.
An illustration of this mode is in \autoref{fig: teaser} (B), showcasing the outcome of fixing the ``observation'' cell. 
Despite being positioned at the end of the notebook, its reference to preceding code and output cells becomes easier. 
The click-to-fix feature ensures the ``observation'' cell remains visible alongside related \yanna{code} and outputs, enhancing content correlation and {understanding}. 
The fixed icon, as seen in \autoref{fig: teaser} (b2), serves as an indicator of the cell's fixed status.
Readers can easily \final{exit} this fixed mode with another click.

In summary, \tool uses these interactions to support exploring the detailed relationships in $R$, allowing readers to delve deeper into specific relationships as needed, without overwhelming the interface with excessive visual information. 



\section{Relationship Specification}

\tool~focuses on proposing a reading-optimized design for \final{presenting relationships, rather than specifying them. 
Additionally, we have developed a simple stand-alone proof-of-concept notebook plugin for specifying relationships}
through direct manipulations.
In this plugin users can select relevant text or code and draw a bounding box or freehand sketch on the output to indicate that selected cells or segments are related.
\final{The plugin then records the relationships specified through users' interactions in a text file}, formatted according to the defined formulation.
Using this relationship file, \tool further visualizes and presents the relationships, as described in~\autoref{sec:system_design}.
Please refer to the supplementary material for more details.


\begin{figure*}[t!]
\textbf  \centering
  \includegraphics[width=1.0\textwidth,alt={This figure illustrates an example comprehension question from the user study and how participants could navigate and use \tool to arrive at the correct answer. Part (A) displays a multiple-choice question: "What operations were performed on TotalBsmtSF before applying the log transformation, and why were these steps taken?" The options include: replacing missing values with the feature's mean to address accuracy issues (A), replacing zero values with the feature's mean to avoid logarithmic transformation issues (B), deleting all records where TotalBsmtSF values were zero as erroneous entries (C), creating a new variable named HasBsmt to distinguish cases where TotalBsmtSF is zero to handle logarithmic transformations (D, the correct answer highlighted in red), and the option "I don't know" (E).
Part (B) contains the corresponding notebook content needed to answer the question, displayed in three sections. The first section (b1) highlights the code performing the log transformation on TotalBsmtSF, specifying that it only applies to non-zero values. The second section (b2) provides textual commentary that the log transformation ignores zero values. The third section (b3) displays the preceding code that creates the variable HasBsmt to distinguish whether TotalBsmtSF is zero. The final section (b4) offers a textual explanation (via hover interaction) confirming that HasBsmt is used to handle zero values, allowing the log transformation to proceed correctly.
Through these steps, participants can logically deduce that the correct answer is D.}]{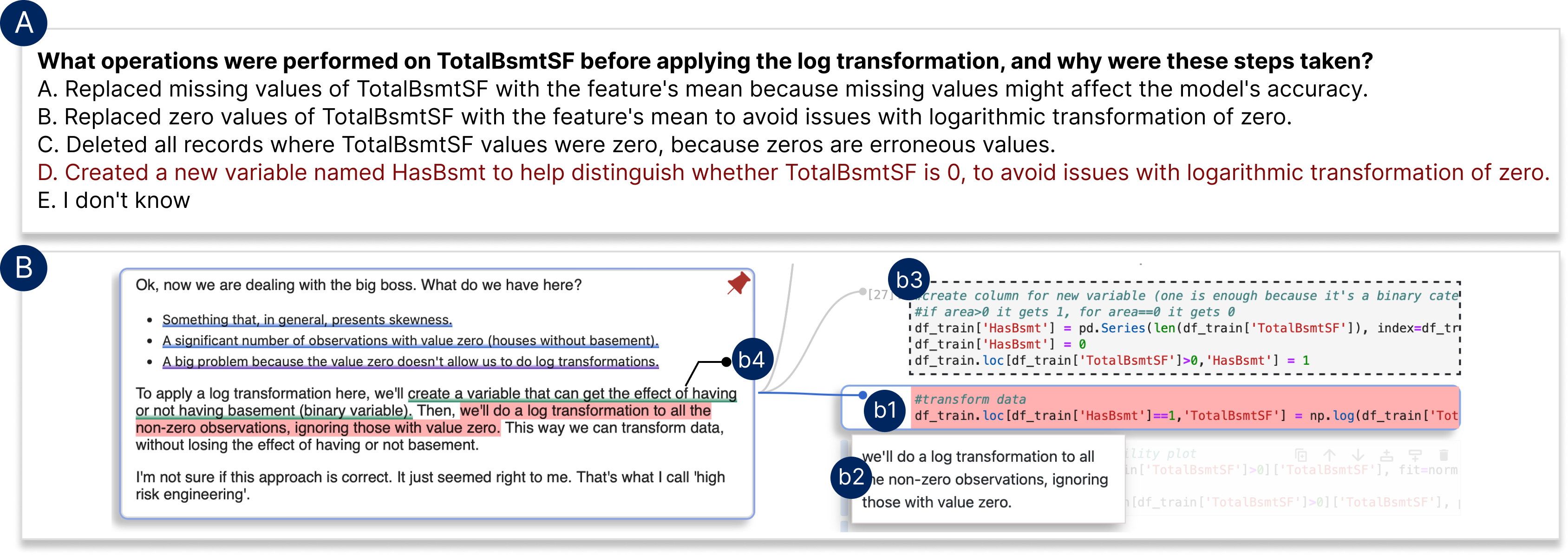}
  \caption{\yanna{An example question from the user study and a potential process to get the correct answer using \tool. (A) displays the question, while (B) provides the relevant information needed to answer it. 
  To get the correct answer, participants could first locate the cell performing the log transformation on \textit{TotalBsmtSF} (b1), \final{and} identify that the log operation excludes values of zero (b2). They would then examine the preceding code (b3) and hover over it to access the textual explanation (b4), revealing that a new variable, \textit{HasBsmt}, was created to distinguish whether \textit{TotalBsmtSF} is zero or not. Based on this process, participants could identify the correct answer as \textit{D}.}}
  \label{fig:example_question}
  \vspace{-1em}
\end{figure*}

\section{Evaluation}

We conducted a within-subject user study with 12 participants to evaluate the usability and effectiveness of \tool in  \final{facilitating identifying and navigating relationships between text, code, and outputs in notebooks}.
Participants were tasked with 
answer\final{ing} questions based on provided notebooks, using both \tool and the traditional Jupyter notebook interface as \final{a} baseline for comparison.
The study setting is detailed in~\autoref{sec:study_setup}, followed by an analysis of the results in~\autoref{sec:results}.

\subsection{Study Setup}
\label{sec:study_setup}

\textbf{Participants.}
We recruited 12 participants (P1-P12, 6 men and 6 women, 22-31 years old, average age 25.1) through social media and word-of-mouth advertisements.
They were two software engineers, nine postgraduate researchers, and one postdoctoral researcher with diverse backgrounds, including visualization, human-computer interaction, database management, programming languages, geographic information systems, and computer vision.
\yanna{Participants reported average familiarity scores of 3.7 for Jupyter Notebook and 4.2 for Python, on a scale from 1 (not at all familiar) to 5 (extremely familiar).}
Each participant received compensation of US \$10/hour for completing the user study.

\textbf{Study materials.}
We first selected two notebooks,~\ie~the \textit{House} notebook\footnote{https://www.kaggle.com/code/pmarcelino/comprehensive-data-exploration-with-python} 
and \textit{Titanic} notebook\footnote{https://www.kaggle.com/code/startupsci/titanic-data-science-solutions},
from the popular Kaggle competitions~\cite{wang2022documentation}.
They have been widely used to evaluate the effectiveness and usability of the computational notebook plugins~\cite{li2023notable, wang2024outlinespark, zheng2022nb2slides}.
Specifically, \textit{House} notebook contains 55 text cells and 32 code cells with 24 output cells, while \textit{Titanic} notebook contains 49 text cells and 52 code cells with 49 output cells.

\revise{We invited two participants from the formative study (FP2 and FP4), each to help identify the relationship set $R$ for one of the notebooks in the format mentioned in \autoref{sec:relationship_formulation}.
We asked them to assume they were the authors of the notebook preparing their content to share their notebook with a wide audience. 
This task involved annotating the relationships between the information they believed would facilitate understanding the notebooks.}
Subsequently, we utilized \tool to display these relationships, and invited FP2 and FP4 to review and, if necessary, refine the relationships. 
Through this iterative process, the \textit{House} notebook contains 56 relationships, while the \textit{Titanic} notebook contains 100 relationships.
The specifics of these relationships, along with their distribution across different types, are reported in the supplementary material.

We designed eight questions for each notebook to encompass eight distinct relationship types.
\yanna{Following the questions design in~\cite{wang2020callisto}, we designed the questions according to the Bloom's taxonomy~\cite{armstrong2010bloom} to assess \final{how well participants could identify and navigate relationships}.
The answer to each question required information from at least two cells, requiring participants to identify \final{and navigate} the relationships between different pieces of information.
\autoref{fig:example_question} presents one example choice-based question, where participants are required to \textit{identify} the operations performed before applying a log transformation and \textit{explain} the rationale behind those steps using information from three cells.
}
Choice-based questions were chosen over open-ended questions to mitigate issues observed during a pilot study with researchers out of this project:
1) Participants spent most of their time retrieving information rather than engaging in deep understanding.
When asked to answer questions using textual descriptions and providing evidence via screenshots, they often copied and pasted answers linked by \tool, without making an effort to \final{read and understand} the content; and
2) Evaluating answers to open-ended questions is challenging, introducing subjectivity and potential bias.
\yanna{Distractor answer items were generated using GPT-4~\cite{openai2023gpt4}, prompted to create plausible yet incorrect options based on the given question and correct answers.}
Details of questions can be found in the supplementary material.

\begin{figure}[t!]
\textbf  \centering
  \includegraphics[width=\linewidth,alt={A series of four box plots comparing task performance metrics between a baseline tool, represented by gray triangles, and InterLink, represented by black triangles. From left to right, the metrics are task accuracy, confidence level, average time spent on tasks, and IES Score. A star above the comparisons indicates a statistically significant difference with a p-value less than 0.05. The first and last box plots, representing accuracy and IES score, show significant improvements when using InterLink over the baseline, as indicated by the stars. The confidence depicts higher medians while time box plots depict lower medians for 'InterLink', suggesting marginally improved average time and confidence levels when using this tool.}]{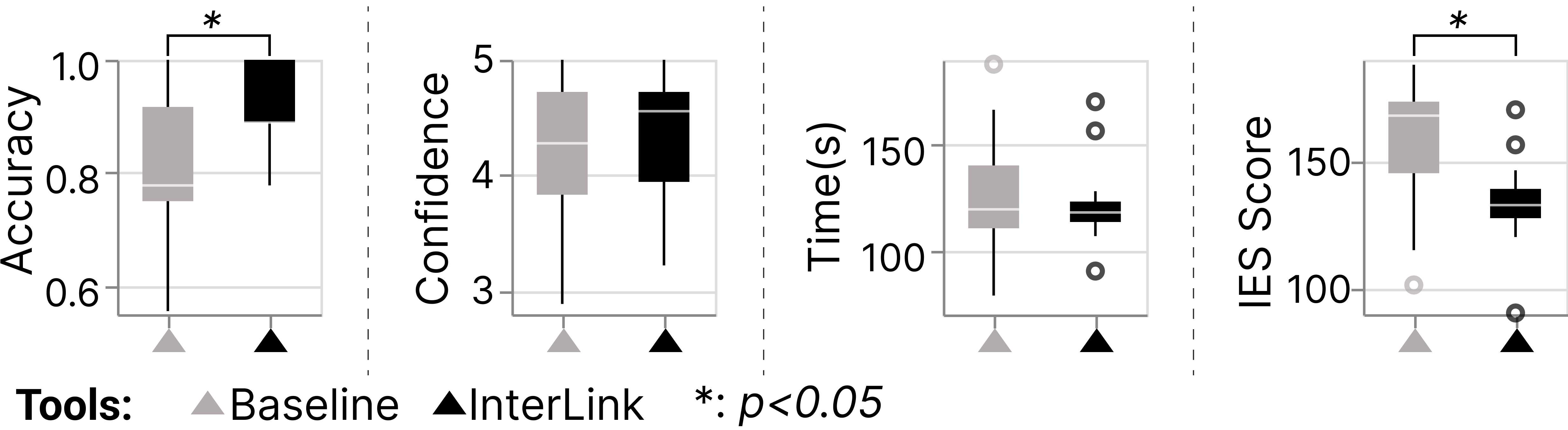}
  \caption{Task performance metrics between baseline and \tool, including task accuracy, confidence level, average time, and IES score, with * indicating $p<0.05$. Overall, participants using \tool demonstrated significantly higher task accuracy and \yanna{lower} IES score, along with marginally improved average time and confidence levels.}
  \Description{A series of four box plots comparing task performance metrics between a baseline tool, represented by gray triangles, and InterLink, represented by black triangles. From left to right, the metrics are task accuracy, confidence level, average time spent on tasks, and IES Score. A star above the comparisons indicates a statistically significant difference with a p-value less than 0.05. The first and last box plots, representing accuracy and IES score, show significant improvements when using InterLink over the baseline, as indicated by the stars. The confidence depicts higher medians while time box plots depict lower medians for 'InterLink', suggesting marginally improved average time and confidence levels when using this tool.}
  \label{fig:task_performance}
  \vspace{-1em}
\end{figure}

\textbf{Tasks.}
In our study, participants were tasked with \final{reading} two provided notebooks and completing a test using two interfaces:~\tool and baseline (\ie~the traditional JupyterLab interface).
In each interface, participants \yanna{had} five minutes to overview the notebooks.
\yanna{They then answered eight questions to \final{assess how well they could identify and navigate relationships}, with the notebook remaining visible to \final{avoid the need for demanding memorization}~\cite{kong2019understanding}.
}
Participants completed tasks using the two interfaces across two distinct notebooks.
The experiment included all possible combinations of the two interfaces and two notebooks. To reduce learning effects, the order of the two interfaces was counterbalanced.

\textbf{Procedure.}
The user study lasted approximately 1.5 hours and was conducted through individual online meetings. 
Initially, we presented the project's background and procedures, followed by obtaining the participants' consent to record the sessions.
Regarding \tool, participants undertook a brief tutorial session to familiarize themselves with \tool's components and functionalities, subsequently practicing with \tool using a sample Kaggle notebook
\footnote{https://www.kaggle.com/code/mohitkr05/spotify-data-visualization}. 
The tutorial session concluded once participants indicated they felt sufficiently familiar to use the tool, typically within 15 minutes.
After that, participants would answer 2 questions related to the sample notebook. This is to help them be more familiar with the tasks.
Participants were required to explain their answers for us to check and reinforce the requirement that answers should be based solely on notebook content instead of personal knowledge.
Following the tutorial, participants \final{completed} the formal tasks of using two interfaces sequentially.
Answers were recorded along with the time taken and participants' self-reported confidence levels (on a five-point Likert Scale).
Upon completing the tasks with each interface, participants filled out the questionnaire assessing the effectiveness of the interface in  \final{identifying and navigating relationships} on a five-point Likert scale in a think-aloud protocol.
The details of the questionnaire are shown in the~\autoref{fig:quantitative_result}, where a rating of 1 means ``strongly disagree'' and a rating of 5 means ``strongly agree''. 
For \tool, participants also completed the System Usability Scale (SUS) questionnaire~\cite{brooke1996sus} to assess its usability. 
The study concluded with a semi-structured interview discussing \tool's advantages and disadvantages and its effect on \final{participants' identification and navigation experience of relationships and their understanding of the notebooks}.

\subsection{Results}
\label{sec:results}

In this section, we report the quantitative results and qualitative user feedback of the user study.

\subsubsection{Quantitative Results}

\label{sec:quantitative_analysis}

\begin{figure}[t!]
\textbf  \centering
  \includegraphics[width=\linewidth,alt={This figure presents user ratings for effectiveness, with statistical significance indicated by double asterisks denoting a p-value less than 0.01. It compares the effectiveness of InterLink against a baseline tool, showing InterLink as superior in all measured aspects through stacked horizontal bar charts for five questions, Q1 through Q5, with colors ranging from dark red to dark blue indicating rating levels from 1 to 5.}]{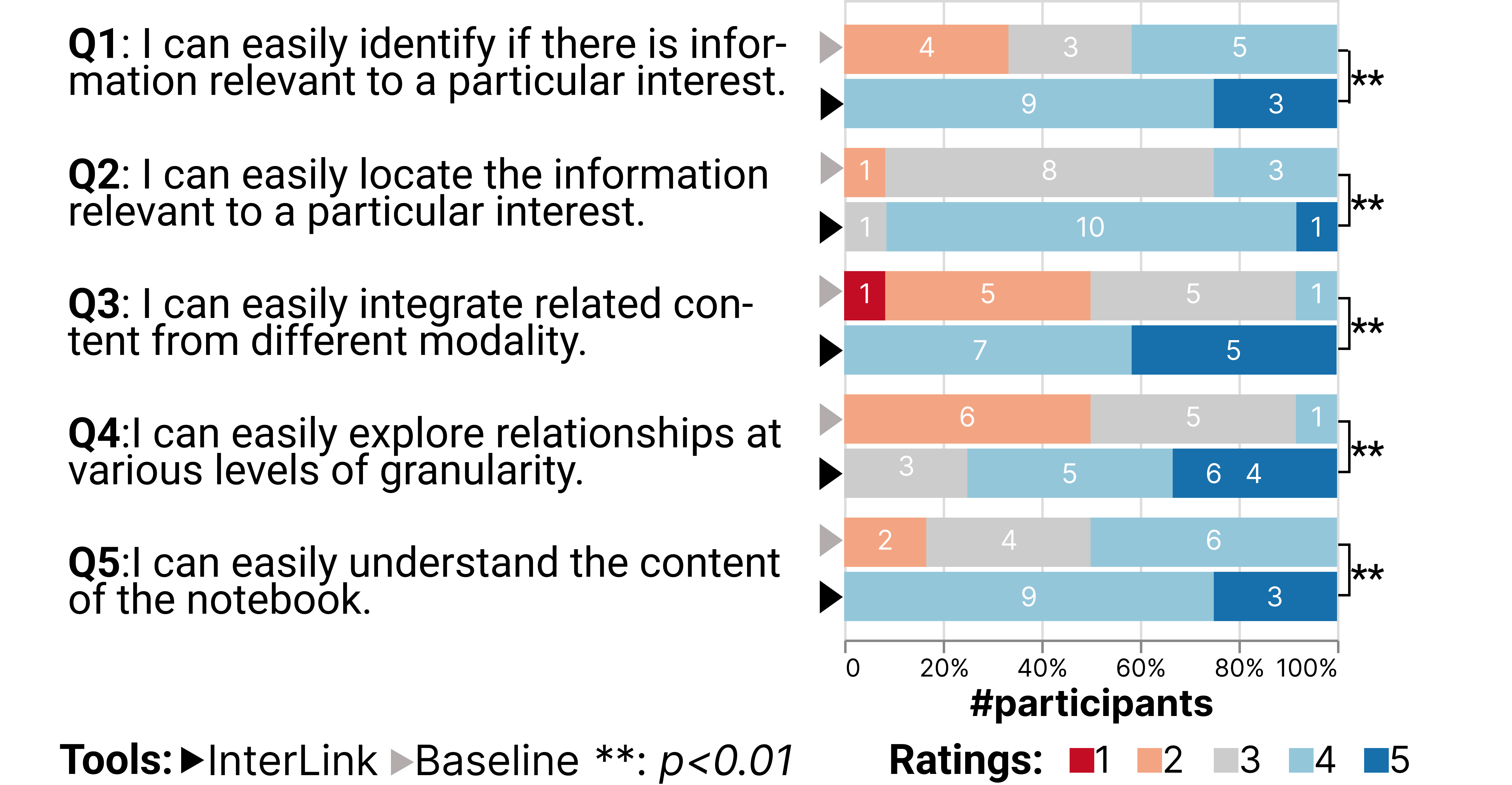}
  \caption{This figure presents user ratings for effectiveness of \tool compared to the baseline, with ** indicating $p<0.01$. It demonstrates that \tool significantly outperforms the baseline across all aspects, with all $p<0.01$.}
  \Description{This figure presents user ratings for effectiveness, with statistical significance indicated by double asterisks denoting a p-value less than 0.01. It compares the effectiveness of InterLink against a baseline tool, showing InterLink as superior in all measured aspects through stacked horizontal bar charts for five questions, Q1 through Q5, with colors ranging from dark red to dark blue indicating rating levels from 1 to 5.}
  \label{fig:quantitative_result}
  \vspace{-1em}
\end{figure}

Next, we describe the quantitative results across three aspects, \ie overall task performance, effectiveness, and usability, with results shown in ~\autoref{fig:task_performance} and ~\autoref{fig:quantitative_result}.

\textbf{Overall performance.}
We measured the task performance of participants using task completion time, task accuracy, confidence levels, and the Inverse Efficiency Score (IES)~\cite{townsend1983ies, zhi2019linking, shi2020calliope}.
Task accuracy was determined by the proportion of correct responses.
Considering that participants may focus on arriving at correct answers while sacrificing time, \final{the} IES was used to balance accuracy and time costs.
It was calculated by dividing the task completion time by the task accuracy.

The results of task performance are shown in~\autoref{fig:task_performance}.
Participants using \tool complete\final{d} tasks \final{13.6\%} more accurately ($0.92\pm0.07$ for \tool and $0.81\pm0.14$ for baseline) and demonstrated a \yanna{lower} IES score ($134\pm19.5$ for \tool and $158\pm28.1$ for baseline), with \yanna{paired t-test} suggesting that the differences are statistically significant (both $p<0.05$).
\tool also showed slight improvements in the average time taken to answer each question (122 seconds for \tool and 128 seconds for baseline) and in self-reported confidence levels (4.3 for \tool and 4.2 for baseline). These differences, however, were not statistically significant.
\revise{Furthermore, participants reported no significant differences in performance when interacting with various notebooks across all aspects (all $p > 0.05$).}

\textbf{Effectiveness.}
\yanna{\autoref{fig:quantitative_result} shows that \tool received significantly higher user ratings for effectiveness in helping \final{users identify and navigate relationships in notebooks} compared to the baseline interface.}
Specifically, \tool significantly \yanna{facilitated participants} to infer the existence of relationships within the notebook (Q1), locate related information more efficiently (Q2), integrate related information for a comprehensive understanding (Q3), and delve into the relationships and related information at various levels of detail (Q4), thereby improving overall notebook understanding (Q5). 
Each of these improvements was statistically significant, with all p-values less than 0.01.
Participants appreciated the effectiveness of \tool in aiding these aspects, with average scores exceeding 4 on a 5-point Likert scale.

\textbf{Usability.}
The usability of \tool, as measured by the System Usability Scale (SUS), indicates a score of 76.5, positioning it as acceptable and better than approximately 80\% of applications~\cite{sauro2016quantifying, bangor2008empirical}. 
Participants generally rated positively framed SUS questions (the odd-numbered ones) above 4 and negatively framed questions (the even-numbered ones) below 2 on average. 
The exception was (Q4), with an average rating of 2.9, indicating that some participants needed technical support in using the system.
More details of the usability ratings are in the supplementary material.

\subsubsection{User Feedback}
\label{sec:qualitative_analysis}
This section presents participants' qualitative feedback about the advantages and limitations of~\tool.

\revise{
\textbf{\tool allows users to focus on \final{understanding by streamlining the retrieval of related contents}.}
Participants generally appreciated how \tool enables \final{them to identify and navigate relationships, making it intuitive and easy to integrate the related contents.}
This \final{reduced their effort spent on} retrieving and recalling related information, allowing them to devote more time \final{to reading and therefore} deepening their understanding. 
For instance, P1 commented, ``\textit{I can spend more time on understanding rather than searching for related information, which is really tedious and annoying}''.
Similarly, P6 noted, ``\textit{I don't need to think about where the related content is anymore—it's right there, and all I need to do is just understanding.}''
Moreover, some participants (P5, P9, and P12) expressed a desire for the tool to be open-sourced and integrated into their existing platforms.
P12 specifically noted, ``\textit{I hope to use this tool in my JupyterLab soon.
Moreover, extending this functionality to other platforms with interconnected contents, such as PowerPoint or Word documents, would be highly valuable.}''
}

\textbf{The relationship visualization of \tool facilitates locating and focusing on related content in computational notebooks.}
All participants recognized that explicit lines and visual cues in \tool significantly ease the process of inferring and locating related information within computational notebooks. 
This feature effectively reduces the need to filter irrelevant information, allowing readers to direct their limited attention to areas that warrant closer examination and avoid unnecessary distractions. 
For example, P1 noted that ``\textit{This design let me have a quick idea of what is related, so I can focus on the critical parts, significantly boosting my productivity}''.
P9 further underscored the utility of such visual aids, noting, ``\textit{such visualizations cohere loosely connected information, enhancing my overall comprehension}''.
Additionally, P2 expressed relief provided by these visual elements, stating, ``\textit{the lines alleviate my concern \final{about} overlooking vital information}''.
Furthermore, participants particularly valued the finer granularity of information provided by visual cues, recognizing its crucial role in clarifying precise and accurate connections across cells. 
This detailed guidance is particularly valuable when the relevant text or \yanna{code} is merely minor parts of a larger cell or in understanding the complex visualizations (as mentioned by P1, P2, P5, P6, P7, P10).
For example, P5 stated that such granularity is critical in ``\textit{enabling a quick grasp of information and significantly reducing the time I spend reading extensive texts and the effort needed to interpret the visualizations}''.
Despite the advantages, P11 voiced a concern about potentially overlooking information not highlighted by visual cues.

\textbf{The hover-to-highlight and click-to-fix interactions are praised for their intuitiveness.}
\tool's interaction design was praised for its intuitiveness, particularly hover-to-highlight and click-to-fix, aligning well with readers' typical information retrieval habits. 
For example, P6 mentioned that ``\textit{
The interaction feels really seamless and helps me know how things are connected in the notebook. 
Especially the part where you hover to highlight things and click to keep them visible – it's super intuitive and just what I'd expect when I'm trying to find and link up information}''.
While P7 appreciated the bidirectional capability for not disrupting the reading flow, stating, ``\textit{The two-way interaction respects my mental model to locate and explore the related information. 
Following the notebook's reading order previously (in \final{the} traditional interface) enforced a forward retrieval of information, since the backward retrieval would disrupt my reading flow}''.

\textbf{The focus mode of \tool facilitates the integration of dispersed information.}
Most participants, except P3, highly praised the focus mode for integrating all related information into a compact, in-situ format on a single page that optimized screen space usage.
This compact presentation was particularly appreciated for reducing cognitive load by minimizing the need to remember and filter unrelated details to find relevant connections, which happens frequently when using the traditional layout (P4, P5, P6, P7).
P4 and P6 further emphasized the benefit of in-situ information presentation in accessing related information while maintaining reading flow from being disrupted by excessive scrolling.
In contrast, P3 thought the focused mode to be redundant, preferring scrolling methods supported by click-to-fix features.
Though useful, some participants complained that the activation of the focus mode through the ``Shift'' key was not inherently intuitive (P1, P7-P10).
This may lead to a reliance on technical assistance to use \tool~(Q4).
Future work should explore alternative intuitive methods that do not clash with existing JupyterLab commands.

\textbf{The side-by-side layout promotes a holistic view and focused attention for \final{identifying and navigating relationships in notebooks}.}
Most participants echoed that the side-by-side layout of \tool, separating text from \yanna{code} and outputs, significantly enhances the \final{identification} and navigation experience \final{of relationships} within computational notebooks. 
P12 stated that ``\textit{the side-by-side \final{layout} is really intuitive and may be the only solution to show the relationship between different \final{elements}}''.
This layout was appreciated for supporting the holistic view of \final{relationships} and enhancing focus.
It promoted a comprehensive grasp of the notebook's structure, with P3 stating, ``\textit{the text on the left serves as a guide, systematically organizing code and output cells, thereby illuminating the purpose of each code block. 
I can quickly get the entire structural composition and the functionality of each block}''.
P5 gave similar comments: ``\textit{The one-to-many relationships in the side-by-side layout make the function and organization of \final{the} \yanna{code} more apparent.
I can easily grasp the notebook's overall structure and the specific roles of different code blocks}''.
Moreover, participants praised that the layout ensured text was always adjacent to \yanna{code} and outputs, thus providing instantaneous access to additional information as needed.
P7 commented, ``\textit{I can grasp both the holistic structure and local details, transitioning between them seamlessly}''.
P6 added, ``\textit{I can easily get information when confusion arises, facilitating the understanding of explanations}''.
Participants also highlighted how this layout augments their focus on specific aspects of computational notebooks.
P2 and P7 underscore the layout's ability to allow a focused understanding of either text or \yanna{code}, thus minimizing distractions and fostering a clearer overview of the notebook.
P6 remarked, ``\textit{This layout really aligns with users' mindset of going through shared notebooks. It starts with a quick summary of the text, \final{and} then lets you dive in deeper if you need to, which really helps understand the \yanna{code}.}''
Despite the overall positive reception, 
a minority of participants mentioned a learning curve when adapting to this new layout (P8, P9, and P11). 
However, they also recognized the potential of the layout to improve \final{reading notebook} with increased familiarity.
For example, P9 mentioned ``\textit{While I am more accustomed to the traditional linear layout due to years of usage, I perceive the new layout as highly beneficial and anticipate its long-term advantages once fully accustomed}''.
\section{Discussion}
This section discusses the lessons learned during our study and the limitations and future work of the research.

\revise{\textbf{Clear relationships \final{help users focus on understanding by streamlining retrieval of related content}.}
\tool~\final{makes computational notebooks easier to understand} by designing a new side-by-side layout with clear visualization and interactions to \final{identify and navigate} relationships between text, \yanna{code}, and outputs. 
This design allows participants to quickly and accurately locate and \final{synthesize} related information, as evidenced by higher task accuracy.
\yanna{
Some participants reported reduced efforts in retrieving relevant content, enabling more focus on \final{integrating and understanding relevant information}.}
Future research should further streamline the burdensome basic tasks of \final{identifying and navigating} relationships of relevant content, enabling users to concentrate more on understanding and higher-level tasks, such as applying and creating~\cite{forehand2010bloom}.}



\textbf{Gaps between current computational notebook practices and the literate programming principle.}
Computational notebooks, designed on the literate programming principle, aim to integrate \yanna{code}, outputs, and explanatory text into a cohesive narrative~\cite{rule2018exploration,wang2022documentation, chattopadhyay2023make}.
However, current practices often adopt a linear layout with unclear relationships between \yanna{code}, outputs, and text~\cite{chattopadhyay2023make}.
Such designs weaken the explanatory role of the text, turning notebooks into loosely connected scripts that diverge from literate programming principles~\cite{rule2018aiding, wagemann2022five}.
To address this, \tool~introduces a side-by-side layout to present relationships more clearly to help readers integrate information. It therefore realigns notebooks with literate programming principles.
\yanna{User ratings suggest that \tool{} \final{helps} participants infer, locate, navigate, and integrate relevant information, and enhance their overall understanding of the notebook.}
We \final{suggest} future research and tool development should further align computational notebooks with literate programming to maximize their potential for clear and effective communication.


\textbf{Limitations and future work.}
Our efforts to make computational notebooks more readable and understandable unveil several future research opportunities.
First, enhancing the system design would be beneficial. 
Participants expressed a desire for more intuitive methods to navigate and manipulate relationships within the notebook environment. 
Future iterations could focus on simplifying non-intuitive key press interactions (P1, P6-P10) and enabling dynamic layout adjustments, such as adjustable proportions of descriptive text, \yanna{code}, and outputs to accommodate diverse user needs more effectively (P3).
Moreover, integrating advanced language models such as GPT-4~\cite{openai2023gpt4} for content summarization was highlighted, especially valuable when dealing with extensive textual information (P11, P12). 
\revise{Considering all types of relationships could also be promising to \final{make notebooks easier to understand}}.
\yanna{Second, our approach relies on predefined relationships that are manually created and require textual description.
Future work could explore description generation techniques~\cite{lin2023inksight} to enhance notebook quality and thus extend the applicability of \tool.
Additionally, advancing technologies for automatic relationship mining~\cite{lin2023dashboard} and developing authoring tools that facilitate easier relationship establishment between different notebook components~\cite{latif2022kori, sultanum2021leveraging} could further enhance the system.
These advancements should address challenges in relationship maintenance, such as cell edits, rearrangements, and re-runs.}
Third, our evaluation contains limitations that warrant future consideration.
One limitation is that participants may not be proficient enough in \tool given the short training time.
Moreover, due to time management of the study, participants were asked to answer the questions rather than \final{carefully read} the entire notebook, which may not fully align with their practice of  \final{reading and understanding notebooks}.
In the future, we hope to conduct long-term and real-world user studies.
\section{Conclusion}

In this work, we introduce \tool, a computational notebook plugin designed to \final{present relationships between text, code, and outputs, thereby making notebooks easier to understand}. 
\tool aims to address the challenges readers face when trying to \final{read and} understand computational notebooks shared by others, which often requires synthesizing text, code, and outputs \final{by identifying and navigating their relationships}. 
Specifically, it features a side-by-side layout that reorganizes notebook elements into two columns and incorporates visual cues and interactions to aid cross-referencing and \final{information integration}.
This reading-optimized design has been shown to improve \final{finding and integrating relevant information for understanding notebooks}, as evidenced by increased accuracy and reduced IES scores in our user study.
We hope our work emphasizes the importance of design considerations in computational notebooks that prioritize not only analysis but also communication.

\begin{acks}
We are grateful to Dongyang Zhong, Lin-Ping Yuan, Liwenhan Xie, Rui Sheng, and Chenyang Zhang for their kind help.
This work is partially supported by the Hong Kong RGC GRF grant 16210722.
\end{acks}

\bibliographystyle{ACM-Reference-Format}
\bibliography{references}

\end{document}